\documentclass{article}

\def\emph#1{{\em #1}}
\def\***#1{{\sc *** #1 ***}}

\let\ROSAT\rosat

\def\mag{\ensuremath{^{\rm m}}}
\def\ergs {ergs$\;$s$^{-1}\,$cm$^{-2}$}
\def\deg{\ensuremath{^{\circ}}}

\usepackage{graphicx}
\usepackage{emulateapj}
\usepackage{timesfonts}

\begin{document}

\submitted{To appear in August 1, 1998 issue of ApJ (vol 502)}

\righthead{VIKHLININ ET AL.}
\lefthead{CATALOG OF ROSAT CLUSTERS}

\title{A CATALOG OF 200 GALAXY CLUSTERS SERENDIPITOUSLY DETECTED IN THE
\ROSAT\/ PSPC POINTED OBSERVATIONS\altaffilmark{1}}
\author{A.\ Vikhlinin\altaffilmark{2}, B.\ R.\ McNamara, W.\ Forman,
C.\ Jones}
\affil{Harvard-Smithsonian Center for Astrophysics, 60 Garden St.,
Cambridge, MA 02138;\\ avikhlinin, bmcnamara, wforman, cjones
@cfa.harvard.edu}

\medskip
\author{H.\ Quintana\altaffilmark{3}}
\affil{Dpto.\ de Astronomia y Astrofisica, Pontificia Universidad Catolica,
Casilla 104, 22 Santiago, Chile}

\medskip
\author{A.\ Hornstrup}
\affil{Danish Space Research Institute, Juliane Maries Vej 30,
2100  Copenhagen O, Denmark}

\altaffiltext{1}{Optical observations reported here were obtained at the
Multiple Mirror Telescope, a joint facility of the Smithsonian Institution
and the University of Arizona, ESO 3.6m and Danish 1.54m telescopes at La
Silla, and the FLWO 1.2m telescope.} 

\altaffiltext{2}{Also Space Research Institute, Moscow, Russia}
\altaffiltext{3}{Presidential Chair in Science}

\begin{abstract}

We present a catalog of 200 clusters of galaxies serendipitously detected in
647 \ROSAT\/ PSPC high Galactic latitude pointings covering 158 square
degrees.  This is one of the largest X-ray selected cluster samples,
comparable in size only to the \ROSAT\/ All-Sky Survey sample of nearby
clusters (Ebeling et al.\ 1997). We detect clusters in the inner
17.5\arcmin\ of the \ROSAT\/ PSPC field of view using the spatial extent of
their X-ray emission.  Fluxes of detected clusters range from
$1.6\times10^{-14}$ to $8\times10^{-12}\,$\ergs\ in the 0.5--2~keV energy
band. X-ray luminosities range from $10^{42}~$erg~s$^{-1}$, corresponding to
very poor groups, to $\sim5\times10^{44}\,$erg~s$^{-1}$, corresponding to
rich clusters. The cluster redshifts range from $z=0.015$ to $z>0.5$. The
catalog lists X-ray fluxes, core-radii, spectroscopic redshifts for 73
clusters and photometric redshifts for the remainder. Our detection method,
optimized for finding extended sources in the presence of source confusion,
is described in detail. Selection effects necessary for a statistical
analysis of the cluster sample are comprehensively studied by Monte-Carlo
simulations.

We have optically confirmed 200 of 223 X-ray sources as clusters of
galaxies.  Of the remaining 23 sources, 18 are likely false detections
arising from blends of unresolved point X-ray sources, and for 5 we have not
obtained deep CCD images.  Above a flux of $2\times10^{-13}\,$\ergs, 98\% of
extended X-ray sources are optically confirmed clusters. The $\log N - \log
S$ relation for clusters derived from our catalog shows excellent agreement
with counts of bright clusters derived from the \emph{Einstein}\/ Extended
Medium Sensitivity Survey (Henry et al.\ 1992) and \ROSAT\/ All-Sky Survey
(Ebeling et al.\ 1997). At fainter fluxes, our $\log N - \log S$ relation
agrees with the smaller-area WARPS survey (Jones et al.\ 1998). Our cluster
counts appear to be systematically higher than those from a 50~deg$^2$
survey of Rosati et al.\ (1998). In particular, at a flux of
$2\times10^{-13}\,$\ergs, we find a surface density of clusters of
$0.57\pm0.07$ per square degree, which is a factor of 1.3 more than found by
Rosati et al. This difference is marginally significant at the $\sim 2$
sigma level. The large area of our survey makes it possible to study the
evolution of the X-ray luminosity function in the high luminosity range
inaccessible with other, smaller area \ROSAT\/ surveys.

\end{abstract}


\section{Introduction}

Clusters of galaxies are among the most important objects for cosmological
studies.  Models of large scale structure formation such as CDM, predict
that the abundance of clusters is determined by the spectrum of primordial
perturbations and cosmological parameters $\Omega$ and $\Lambda$.
Observations of clusters at different redshifts can be used to constrain
these parameters (e.g., White \& Rees 1978, Kaiser 1986, White, Efstathiou,
\& Frenk 1993, Henry \& Arnaud 1991, Viana \& Liddle 1996, Henry 1997).
Following a different approach, observations of the Sunyaev-Zel'dovich
effect (Sunyaev \& Zel'dovich 1972) in a large sample of distant clusters
can be used for a direct measurement of the distance to these clusters, and
thus provide the values of $H_0$ (e.g., Birkinshaw, Hughes, \& Arnaud 1991)
and~$q_0$.

Up until the present, the largest samples of distant clusters resulted from
optical surveys that searched for enhancements in the surface density of
galaxies (e.g., Postman et al.\ 1996). This method suffers seriously from
projection effects (e.g., van Haarlem et al.\ 1997). Distant clusters found
by such techniques as galaxy concentrations around distant radio sources
(Dickinson 1996) or ``dark'' lenses (Hattori et al.\ 1997) cannot be
considered as statistical samples.  Of all methods for detecting distant
clusters, X-ray surveys are the least sensitive to projection, because the
X-ray emission is proportional to the square of the density of the hot gas,
which must be compressed in a deep potential well for us to detect it. It is
noteworthy that unlike optical, X-ray surveys have the possibility of
finding interesting objects such as ``fossil'' clusters in which almost all
galaxies have merged to form a cD galaxy (Ponman et al.\ 1994), and
hypothetical ``failed'' clusters in which galaxy formation was suppressed
(Tucker et al.\ 1995). To date, the largest published sample of distant
X-ray selected clusters is that from the \emph{Einstein}\/ Extended Medium
Sensitivity Survey (EMSS; Goia et al.\ 1990, Stocke et al.\ 1991). However,
because of the relatively high flux limit, the EMSS sample contains only 6
clusters at $z>0.5$.

 Finding clusters in X-rays is complicated by their rarity among other types
of sources.  A comparison of the $\log N - \log S$ relations for all sources
(Hasinger et al.\ 1993a) and clusters (this work) shows that at a flux of
$10^{-14}\,$\ergs\ in the 0.5--2~keV band, clusters comprise not more than
10--20\% of the total source population. The large amount of optical
identification work needed for cluster selection can be greatly reduced if
they are searched for among spatially extended X-ray sources. Even at $z=1$,
a rich cluster with a core-radius of 250~kpc has an angular radius of
$>20\arcsec$, which still can be resolved with the \ROSAT\/ PSPC on-axis.
Detection of extended sources requires new analysis techniques.  Even if the
spatial extent is not used for cluster selection, special detection
techniques are needed because clusters at $z\approx 0.2-0.3$ are 3--4 times
broader than the \ROSAT\/ PSPC point spread function.

The idea of selecting distant cluster samples from various \ROSAT\/ surveys
was pursued by different groups in the past few years.  Rosati et al.\
(1995, 1998) searched for clusters in long exposure ($>15$~ksec) \ROSAT\/
PSPC pointed observations with a total area of 50~deg$^{2}$, using optical
identifications of all extended X-ray sources found by wavelet transform
analysis. Their sample consists at present of 70 clusters.  The Wide Angle
\ROSAT\/ Pointed Survey (WARPS, Scharf et al.\ 1997, Jones et al.\ 1998)
uses the Voronoi Tessellation and Percolation technique to detect both
point-like and extended sources, followed by optical identifications of all
sources. The WARPS cluster sample consists at present of 46 clusters found
in \ROSAT\/ pointings with exposures $>8$~ksec, covering 16.2~deg$^{2}$.  A
small sample of 15 clusters at $0.3<z<0.7$ was identified by the SHARC
survey (Collins et al.\ 1997). The RIXOS cluster sample (Castander et al.\
1995) consists of 13 clusters, detected using a technique which was
optimized for point sources. Their results on cluster evolution appear to
contradict other \ROSAT\/ surveys (Collins et al.\ 1997), probably
because the point source detection algorithm had a low efficiency for
detecting extended cluster emission.  Finally, important information about
the surface density of clusters at very low fluxes is provided by several
very deep \ROSAT\/ pointings in which complete optical identifications are
performed (e.g.\ McHardy et el.\ 1997).  Note that because of the small
area, none of the aforementioned surveys is able to study the luminosity
function of distant clusters above $3\times10^{44}$~ergs~s$^{-1}$, where the
deficit of high redshift EMSS clusters was reported (Henry et al.\ 1992).

In this paper, we present a sample of distant clusters selected from 647
\ROSAT\/ PSPC observations of high Galactic latitude targets, covering a
solid angle of 158 square degrees, a factor of three larger than the largest
of the other \ROSAT\/ surveys. The source catalog includes 200 optically
confirmed clusters, and thus is one of the largest X-ray selected samples,
comparable in size only to the \ROSAT\/ All-Sky Survey sample of nearby
clusters (Ebeling et al.\ 1997).  We detect cluster candidates as extended
X-ray sources using the wavelet decomposition technique described in this
paper and Maximum Likelihood fitting of the surface brightness distributions
to determine the significance of the source extent. We then identify only
significantly extended sources with optical follow-up observations. Optical
observations confirm that 90\% of our sources are indeed clusters of
galaxies.  Various selection effects such as the fraction of clusters which
remain unresolved or undetected, are studied using extensive Monte-Carlo
simulations. Comparison of the $\log N - \log S$ relation for clusters
derived from our and other \ROSAT\/ surveys shows that our cluster counts at
the bright end are in excellent agreement with those from the \ROSAT\/
All-Sky Survey sample of Ebeling et al.\ (1997). At a flux of
$2\times10^{-13}\,$\ergs, our $\log N - \log S$ relation agrees well with
the WARPS survey (Jones et al.\ 1998), but is somewhat higher than that
found by Rosati et al.\ (1998).

Cluster size and flux estimates throughout the paper use
$H_0=50$~km~s$^{-1}$~Mpc$^{-1}$ and $q_0=0.5$. All X-ray fluxes and
luminosities are reported in the 0.5--2~keV energy band.

\section{X-ray data}\label{sec:bg}

 We analyzed only \ROSAT\/ PSPC pointings at high Galactic latitudes,
$|b|>30\deg$, and low absorption, $N_H<6\times10^{20}\,$cm$^{-2}$, excluding
the 10\deg\ radius regions around the LMC and SMC.  Low Galactic latitude
fields were not used because the absorption is large and nonuniform in these
regions, and because a high density of stars complicates optical
identifications. We also excluded observations of extended targets, such as
known clusters of galaxies, nearby galaxies, SNRs, and star clusters.  As
the only exception, we included the 2146+0413 pointing (\ROSAT\/ sequences
800150 and 800150a01) which was an X-ray follow-up of clusters selected
optically in a blank field.

All individual \ROSAT\/ sequences with listed exposures longer than 2~ksec,
meeting the above criteria and publicly available by April 1996, were
extracted from the data archive at GSFC. Using S.~Snowden's software, we
cleaned the data excluding high background intervals. We also generated
exposure maps using R4--R7 detector maps (energy range 0.5--2~keV) weighted
according to the average PSPC background spectrum. Multiple observations of
the same target were merged.  Observations with cleaned exposures
$<1.5$~ksec were discarded. The final dataset consists of 647 fields,
schematically shown in Galactic coordinates in Fig~\ref{fig:fieldsgal}. We
used only hard band images, 0.6--2~keV, which increases the sensitivity of
cluster detection given that the spectrum of the \ROSAT\/ background is much
softer than that of a typical cluster. This energy band is slightly
different from that used for the exposure map generation, but this
discrepancy results only in a very small, $<2\%$, error in the vignetting
correction in the inner region of the field of view where clusters are
detected. To oversample the PSF adequately, an image pixel size of 5\arcsec\
was chosen.

\bigskip
\centerline{\includegraphics[width=3.5in]{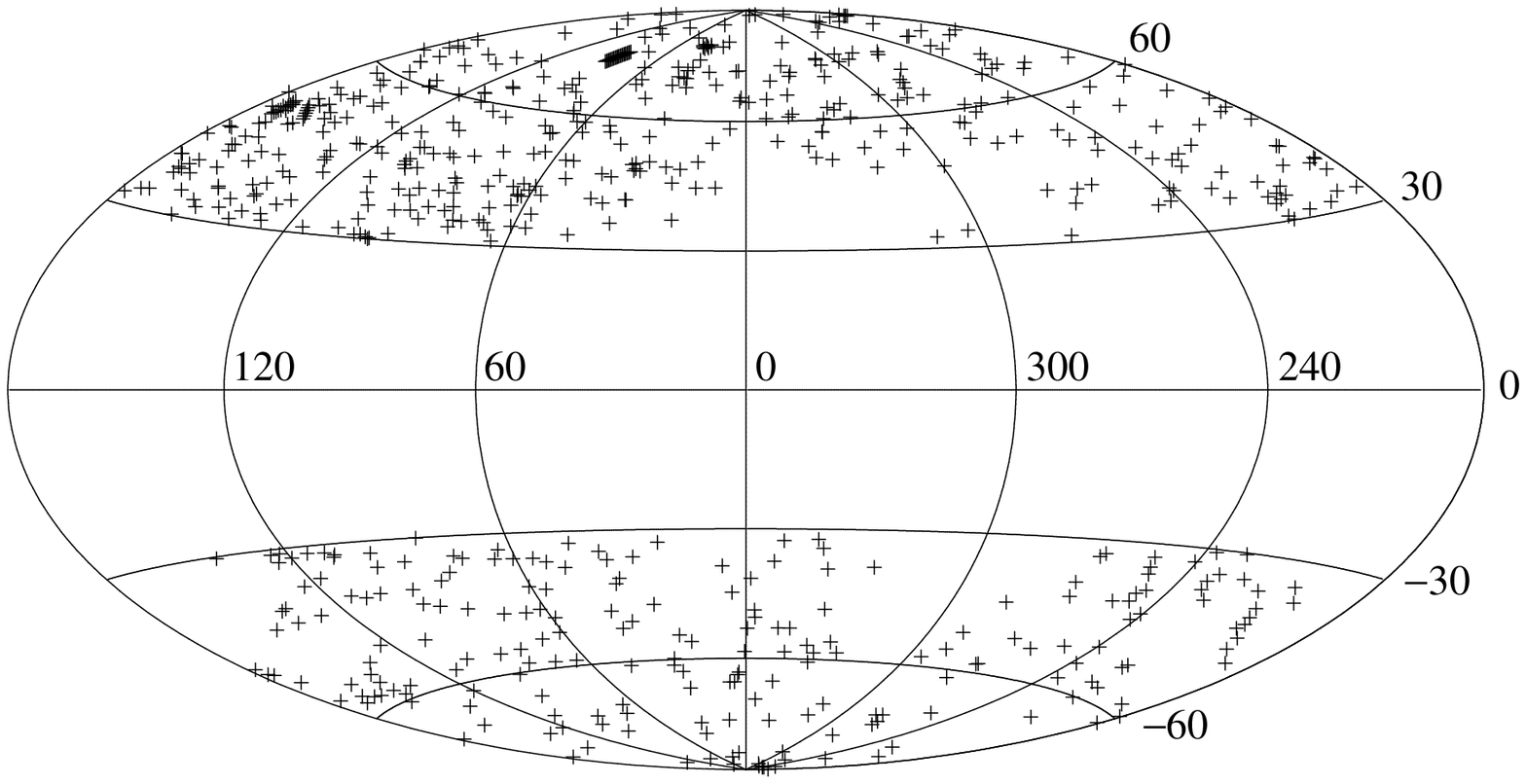}}
\figcaption{The distribution of  \ROSAT\/ pointings in Galactic
coordinates. The higher density in the Northern hemisphere is caused by a
preferential choice of Northern objects as \ROSAT\/ targets.
\label{fig:fieldsgal}}
\medskip

Our next step was to calculate the background map for each observation.  The
\ROSAT\/ PSPC background cannot be modeled simply using the exposure map
template because of the non-uniformity of the cosmic X-ray background, the
presence of scattered solar X-rays, and the wings of the PSF around bright
sources. The angular correlation function of the XRB (Vikhlinin \& Forman
1995, Soltan et al.\ 1996) predicts $\approx 10\%$ brightness fluctuations
on a $10\arcmin$ scale. If not modeled, such background variations can cause
errors in measured cluster fluxes. Since the best approximation to the
background in each field is a smoothed source-subtracted image, we created
background maps as follows.  We first divided an image by its exposure map
to remove the imprint of the PSPC window support structure.  Using the
wavelet decomposition technique (\S\ref{sec:wd}), we subtracted all sources
with characteristic sizes $\leq 3\arcmin$ in radius and smoothed the
cleaned image with a $\sigma=6\arcmin$ Gaussian. The background map was
finally obtained as a product of the smoothed image and the exposure map.

\section{Detection of extended sources}

\subsection{General Considerations}

Finding clusters in \ROSAT\/ PSPC images requires detection of sources of
widely different angular size ranging from approximately the FWHM of the
PSF, $\sim 25\arcsec$, to several arcminutes. Any algorithm for finding
spatially extended sources solves two tasks: A) source detection, i.e.\
identifying regions where the surface brightness significantly exceeds that
of the background, and B) determining extent, i.e.\ deciding whether the
detected source is significantly broader than the point spread function. The
two-stage nature of extended source detection is not usually emphasized, but
can be seen in practice. Rosati et al.\ (1995) convolved images with wavelet
kernels of varying scale to find sources and then derived the source extent
from wavelet amplitudes. Scharf et al.\ (1997) used Voronoi Tessellation and
Percolation (VTP) to find regions with enhanced surface brightness and then
derived the source extent from the measured area and flux. Each of these
methods has advantages for certain tasks. For example, VTP can find extended
sources regardless of their shape. However, none of these methods is optimal
for both parts of the problem.  Obviously, the best sensitivity can be
achieved if, at each stage, one uses a separate algorithm optimized for its
task. We show below that our method of detecting sources using wavelets and
determining source extent by Maximum Likelihood fitting is theoretically
close to optimum for finding regularly-shaped clusters.

The optimal method of source detection is matched filtering (e.g.\ Pratt
1978). For faint sources, the filter is close in shape to the sources
themselves, and any filter with a shape close to the matched one performs
almost equally well (Press et al.\ 1992). Our wavelet detection method uses
filters which approximate Gaussians with $\sigma=1,2,4,\ldots$ pixels. Since
these filters span a range of sizes, nearly optimal detection is achieved
for circular sources of any size. With an axially-symmetric filter, it is
possible to miss very irregular sources. However, most clusters are
relatively regular (Jones \& Forman 1998) for detection purposes. Also, this
shortcoming is clearly outweighed by the merits of the wavelet method, such
as optimal detection of sources with regular shape, complete background
subtraction and elimination of the influence of point sources. We discuss
these issues below in detail.

Consider now the optimal method to discriminate between extended and point
sources. Cluster radial surface brightness profiles can be described by the
so called $\beta$-model, $I(r,r_c)=I_0\,(1+r^2/r_c^2)^{-3\beta+0.5}$ (e.g.\
Cavaliere \& Fusco-Femiano 1976). Therefore, to discriminate between a
cluster and a point source, we should determine whether $I(r,r_c)$ with core
radius $r_c>0$ describes the data better than a $\delta$-function, that is,
$I(r,r_c)$ with $r_c=0$.  According to the \emph{Neyman-Pearson Lemma}
(e.g.\ Martin 1971), the most sensitive test for this problem is the change
in the value of the likelihood function between the best-fit value of $r_c$
and $r_c=0$.  Maximum Likelihood fitting may not be the best method for
finding clusters with arbitrary shape, but theoretically it is the best one
for the vast majority of clusters having regular shape.

Based on the considerations above, we implemented an algorithm for detection
of extended sources which uses our own variant of wavelet transform
analysis, wavelet decomposition, to find all sources even in the presence of
source confusion and Maximum Likelihood fitting of $\beta$-models to
determine whether each source is extended. Each step is discussed below in
detail.

\subsection{Wavelet Detection of Cluster Candidates}\label{sec:wd}

Cluster detection in the \ROSAT\/ PSPC images is complicated by the varying
background and confusion with point sources located in the vicinity of
clusters. The wavelet transform is well-suited to overcome these
difficulties. We briefly outline the relevant properties of the wavelet
transform and then describe our particular implementation. 

\subsubsection{General Properties of the Wavelet Transform}

The basic idea of the wavelet transform applied to astronomical images
(e.g.\ Grebenev et al.\ 1995 and references therein) is a convolution with a
kernel which consists of a positive core and an outer negative ring, so that
the integral of the kernel over the $x,y$ plane is zero. The convolution
with such kernels allows complete background subtraction and isolation of
structures of particular angular size.  This can be shown using a kernel
which is the difference of two Gaussians:
\begin{equation}\label{eq:gausswv}\label{eq:wdfamily}
W(r)\;=\; \frac{\exp(-r^2/2a^2)}{2\pi a^2} \; - \;
\frac{\exp(-r^2/2b^2)}{2\pi b^2},
\end{equation}
where $b=2a$. The convolution of this kernel with any linear function
$s(x,y)=ax+by+c$ is zero. Therefore, any slowly varying background which can
be locally approximated by a linear function is subtracted by a convolution
with this kernel. To demonstrate the ability of wavelets to reveal
structures with a given size, consider the convolution of the wavelet kernel
with a Gaussian $\exp(-r^2/2\sigma^2)$. The convolution amplitude achieves
its maximum when $\sigma=a\sqrt{2}$ but rapidly falls to $1/2$ of the
maximum for $\sigma=a/2$ and $\sigma=4a$. These properties of the wavelet
transform are used for source detection (e.g., Damiani et al.\ 1997). In
most applications, an image is convolved with a family of kernels of the
same functional form while varying its scale ($a$ in eq.~\ref{eq:gausswv}).
Sources are detected as significant local maxima in the convolved images.
Information about the source angular extent can be derived from the wavelet
transform values at different scales. This simple approach works well for
detection of isolated sources, but fails if another bright source is located
nearby, as is shown in Fig.~\ref{fig:wvdecomp}a,b. A point source with a
flux four times that of the cluster is located at $2/3$ core-radii from the
cluster center (\emph{a}). The image is convolved with the wavelet
kernels (eq.\ref{eq:gausswv}) of scale $a=1,2,4,\ldots,32$ pixels
(\emph{b}). At each scale, the point source dominates the
convolution, and the cluster remains undetected. A different kind of
complication for a simple wavelet analysis is caused by compact groups of
point sources. Convolved with a wide kernel, such groups appear as a single
extended source, resulting in false cluster detections. Neither of these
problems can be overcome by using a different symmetric wavelet kernel with
compact support (Strag \& Nguyen 1995). However, they can be overcome using
the idea employed in the CLEAN algorithm commonly applied in radio astronomy
(H\"ogbom 1974): point sources are detected first and subtracted from the
image before the detection of extended sources. Below we describe our
algorithm, which we call wavelet decomposition, which combines this approach
with wavelet transform analysis.

\subsubsection{Wavelet Decomposition}

The family of wavelet kernels we use is given by eq.~\ref{eq:wdfamily}, in
which we use several combinations of $a$ and $b$ which we call scales.  At
scale 1, the positive component in eq.\ref{eq:wdfamily} is a
$\delta$-function ($a=0$) and $b=1$~pixel.  At scale 2, $a=1$ and $b=2$
pixels, at scale 3, $a=2$ and $b=4$ pixels and so on. At the largest scale
$n$, the kernel is a single, positive Gaussian with $a=2^{n-1}$ pixels.  How
close is this family of kernels to the optimal filter for detecting sources
with the $\beta$-model surface brightness profiles? Numerical calculations
show that in at least one of the scales, the signal-to-noise ratio exceeds
80\% of the maximum value corresponding to the optimal filter --- the
$\beta$-model itself --- for $0.55<\beta<0.8$.

The described family of wavelet kernels has the advantage of an easy and
linear back-transformation. The original image $z(x,y)$ is given by
\begin{equation}
z(x,y)= \sum_{j=1}^n w_j(x,y),
\end{equation}
where $w_j(x,y)$ is the convolution with the kernel of scale $j$. An
important interpretation of this wavelet transform follows from this
equation: it provides a decomposition of an image into a sum of components
of different characteristic sizes. With this interpretation, we construct
the following iterative scheme to remove the effect of point sources.

We convolved the image with a kernel of the smallest scale, estimated the
detection threshold as described below, and cleaned the image of noise.  The
convolved image values were preserved in those regions where the brightness
exceeded $1/2$ of the detection threshold and which contained at least one
maximum above the detection threshold. The remaining image was set to zero.
We subtracted this cleaned image from the input image to remove the sources
that have been detected at this step, and repeated the convolution and
cleaning procedure iteratively until no more sources were detected at this
scale. We also added cleaned images obtained at each iteration to produce a
composite image of significant sources detected at this scale. We then moved
to the next scale, at which the input image was set to the original image
minus everything detected at the first scale.  The iterations were stoped at
scale 6, for which $a=80\arcsec$ and $b=160\arcsec$ and detected sources
have typical full widths of 3\arcmin--4\arcmin.

\begin{figure*}[htb]
\vspace*{-3ex}
\mbox{}\hfill \includegraphics[width=3.25in]{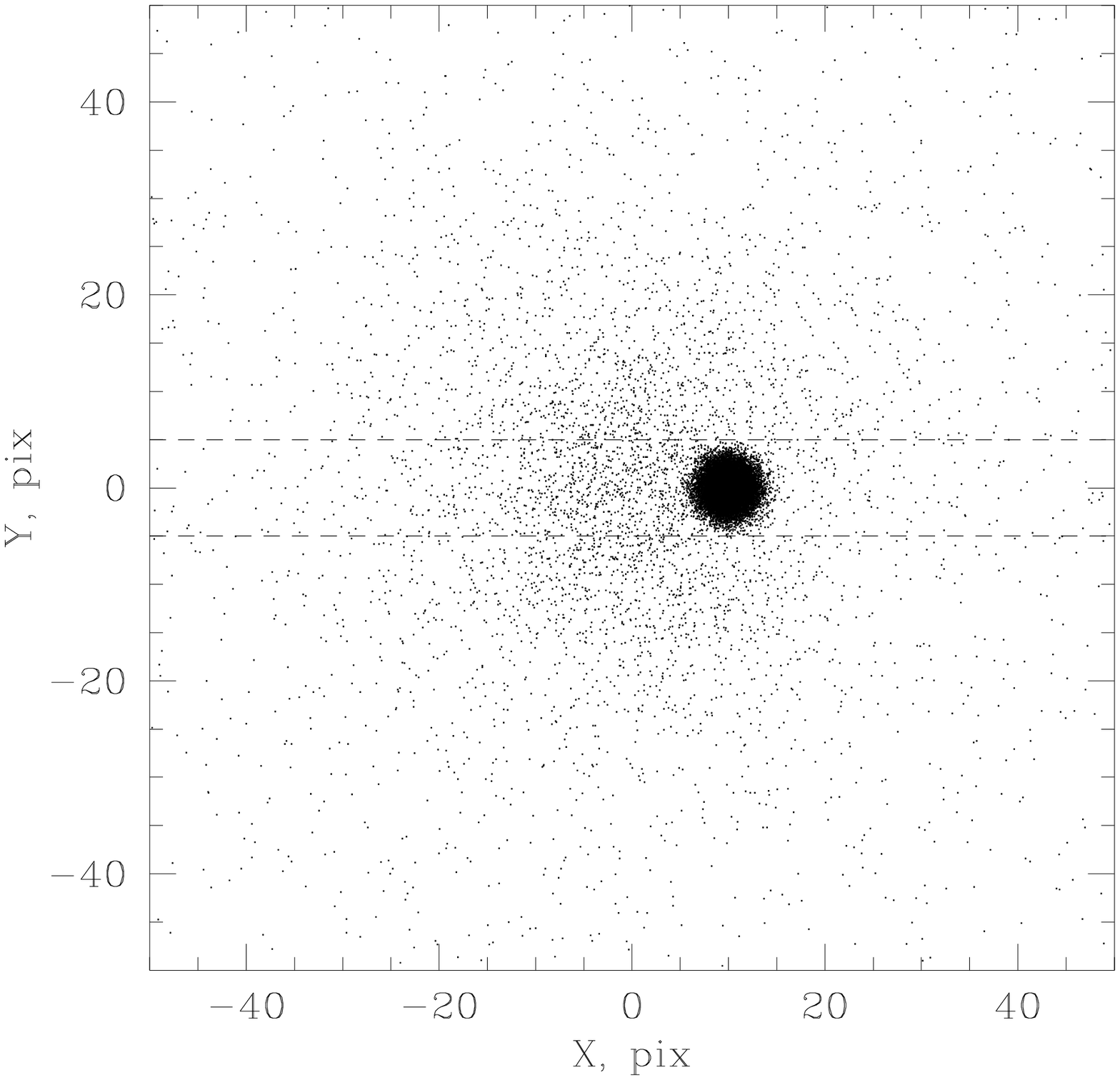} \hfill
\includegraphics[width=3.25in]{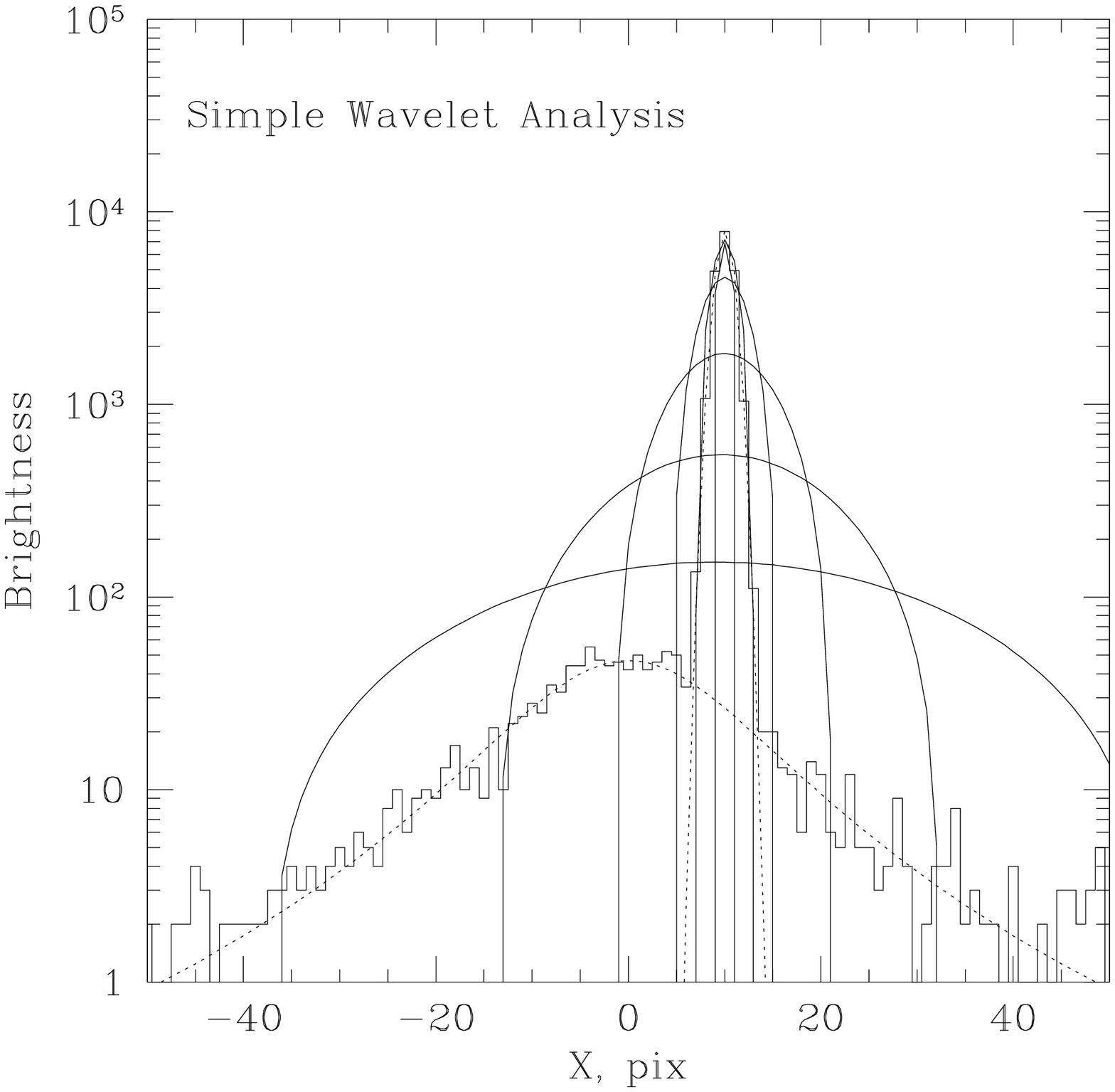} \hfill \mbox{}\par
\vskip -2.5ex
\mbox{}\hfill \includegraphics[width=3.25in]{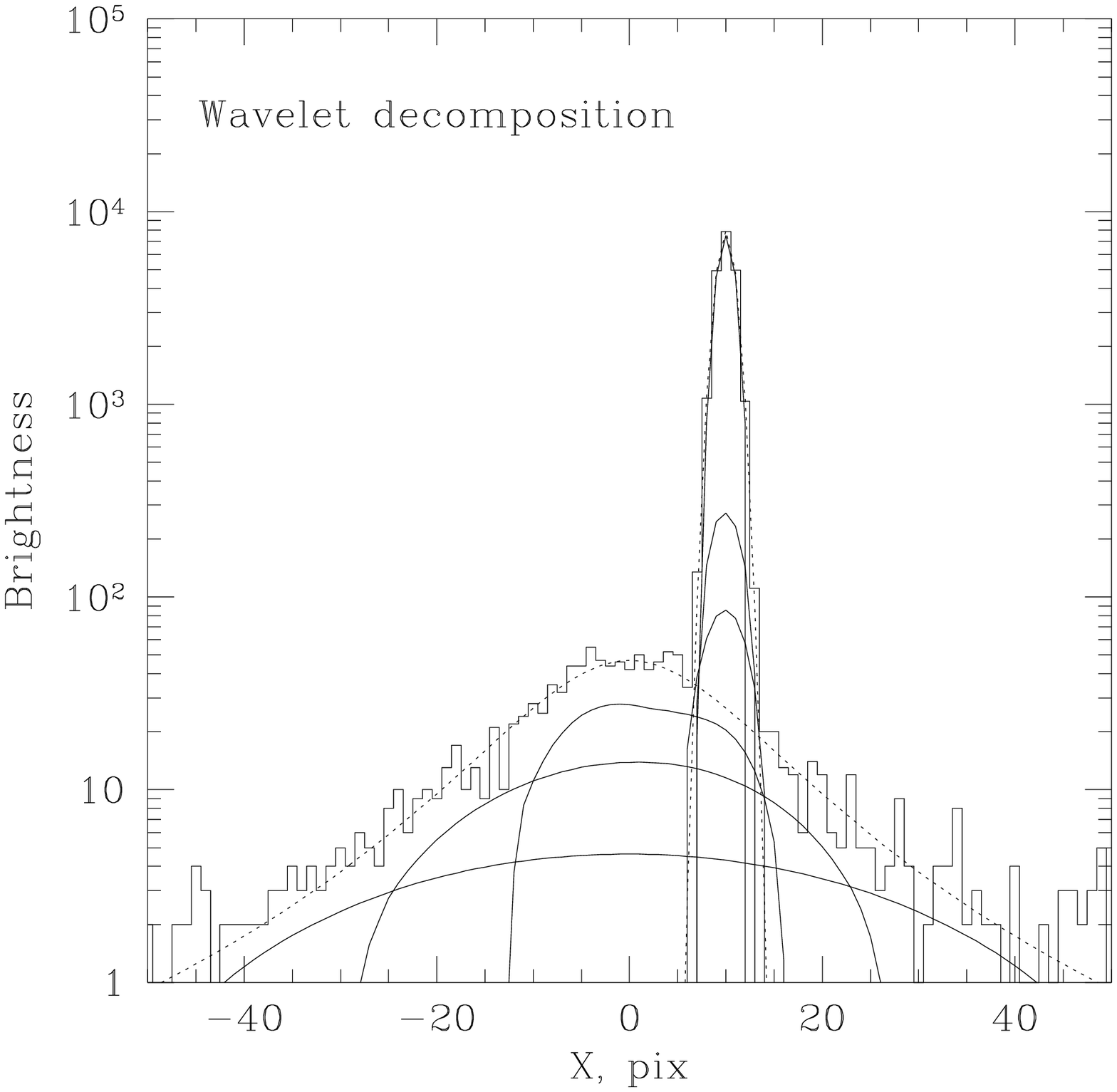} \hfill
\includegraphics[width=3.25in]{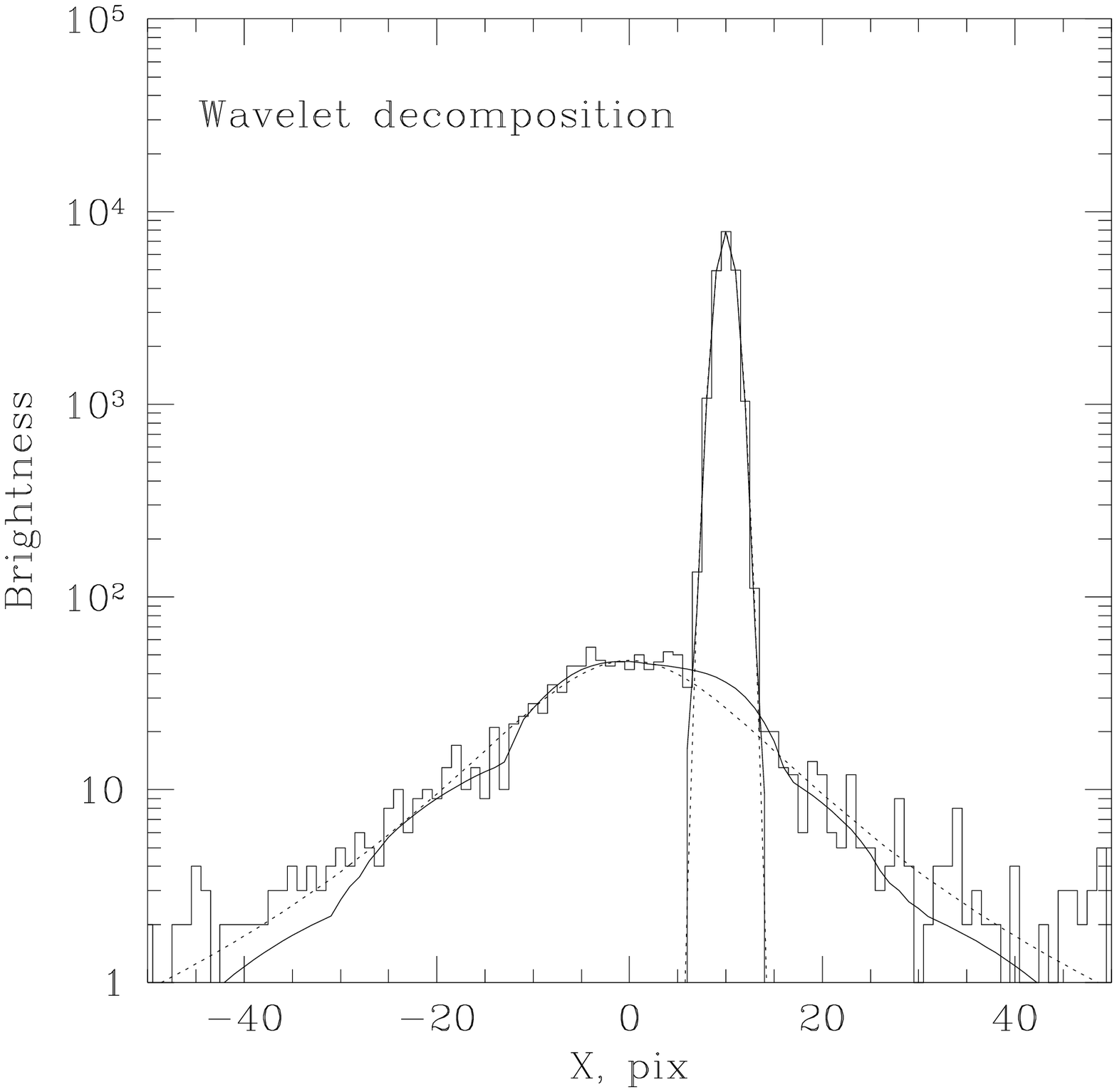} \hfill\mbox{}
\vskip -2.5ex
\caption{\footnotesize Advantage of the wavelet decomposition algorithm. 
A bright point source is located in the vicinity of a cluster \emph{(a)}.
Dashed lines show the strip in which brightness profiles (panels
\emph{b--d}) were extracted. Panel \emph{(b)}\/ shows the result of
convolution of this image with wavelet kernels (eq.\ref{eq:gausswv}) with
the scale $a=1,2,4,\ldots,32$ pixels. The data profile is shown by the solid
histogram, and the profiles of convolved images by solid lines. At all
scales, the convolution is dominated by the point source and there is no
separate peak corresponding to the cluster. Therefore, the cluster remains
undetected by this simple analysis. Our method \emph{(c)}\/ provides a
decomposition of the original image into components with the characteristic
size 1, 2, 4, $\ldots$, 32 pixels.  Small-scale components model the point
source. The cluster becomes apparent and well-separated from the point
source at large scales. The sum of the three smallest and three largest
scales of the wavelet decomposition provide almost perfect decomposition of
the raw image into its original components \emph{(d)}.}
\label{fig:wvdecomp}
\vskip -1.5ex
\end{figure*}

The bottom panels of Fig.~\ref{fig:wvdecomp} illustrate this procedure. The
smallest wavelet kernel is insensitive to the broad cluster emission and
detects only the point source. When iterations at scale 1 are completed,
$\sim 90\%$ of the point source flux has been subtracted. Subtraction of the
point source continues at scales 2 and 3, while the cluster remains
undetected because it is broader than the analyzing kernel. The point source
is almost completely subtracted at small scales and does not interfere with
cluster detection at scales 4--6. The result of these iterations is a set of
images containing statistically significant structures detected at each
scale, whose characteristic size corresponds to the width of the analyzing
kernel at this scale.  Therefore, to separate point and extended sources,
one can combine small and large scales, respectively. As
Fig.~\ref{fig:wvdecomp}d shows, the sum of scales 1--3 and 4--6 provides
almost perfect separation of the original image into the point source and
the cluster.

It is important to choose the correct detection thresholds. Although several
analytic methods of deriving detection thresholds for the wavelet transform
were suggested (Starck \& Pierre 1998 and references therein), we determined
them through Monte-Carlo simulations. We simulated $512\times512$ images
consisting of a flat Poisson background and convolved them with the wavelet
kernels. The distribution of the local maxima in the convolved images was
used to define the detection threshold. We set this threshold at that value
above which one expects to find on average $1/3$ local maximum per simulated
background image per scale in the absence of real sources, so that in the
combined scales 4--6 we expect one false detection per image. Thus defined,
detection thresholds correspond to a formal significance of $\approx
4.5\sigma$.  Detection thresholds were tabulated for a grid of simulated
background intensities. In the analysis of real images, we estimated the
local background and found the detection threshold by interpolation over the
precalculated grid.  Detection thresholds were deliberately set low,
allowing approximately 600 false detections in the entire survey, since our
goal at this step was to detect all possible candidates for the subsequent
Maximum Likelihood fitting, where the final decision about source
significance and extent is made.

As a result of the wavelet decomposition, we obtain six images which contain
detected sources of characteristic size (FWHM) approximately $7\arcsec,
15\arcsec, 30\arcsec, 60\arcsec, 120\arcsec, 240\arcsec$ (scales 1 through
6). We use these images to select candidate extended sources for subsequent
modeling.  Since the FWHM of the PSF is 25\arcsec\ on-axis, most point
sources are detected on scales 1--3 and are absent at scales 4--6. On the
other hand, a distant cluster with core radius of 250~kpc at $z=0.5$ has an
angular radius of 35\arcsec\ (equivalent to $\sim 70\arcsec$ FWHM) and hence
is detected at scales 4--6, to which point sources do not contribute. Even
clusters with smaller core radii, $\sim10\arcsec$, would be detected at
scale 4, because their surface brightness profiles become broader than $\sim
30\arcsec$ FWHM when blurred by the PSF.  Therefore, cluster candidates can
be selected as sources detected at scale 4 or higher.  Some point sources,
especially those at large off-axis angles where the angular resolution
degrades, are detected at scale 4. This shows that our cluster candidate
selection based on the wavelet decomposition is lenient, and we are unlikely
to miss any real clusters at this step. The next step is the Maximum
Likelihood fitting of selected candidate extended sources to determine the
significance of their extent and existence, which will be used for the final
cluster selection.

\subsection{Maximum Likelihood Fitting of Sources}\label{sec:ML}

\subsubsection{Isolated Clusters}

The procedure is straightforward for isolated extended sources. The photon
image is fit by a model which consists of the $\beta$-model convolved with
the PSF. Source position, core-radius, and total flux are free parameters,
while $\beta$ is fixed at a value of $2/3$. The model also includes the
fixed background taken from the map calculated as described in
\S\ref{sec:bg}.  The PSF is calculated at the appropriate off-axis angle for a
typical source spectrum in the 0.6--2 keV energy band (Hasinger et al.\
1993b).  The best fit parameters are found by minimizing $-2\ln L$ (Cash
1979):
\begin{equation}
-2\ln L \; = \; -2 \sum \left(d_{ij}\ln m_{ij} - m_{ij}\right),
\end{equation}
where $d_{ij}$ and $m_{ij}$ are the number of photons in the data and the
model in pixel $(i,j)$, respectively, and the sum is over all pixels in the
fitted region. Note that $m_{ij}$ includes background, so $-2\ln L$ is
defined even if the source flux is set to zero. Along with best-fit
parameters we determine the formal significances of source existence and
extent.  The significance of source existence is found from the change in
$-2\ln L$ resulting from fixing the source flux at zero (Cash 1979).
Similarly, the significance of the source extent is found by fixing the
core-radius at zero and re-fitting the source position and flux.

\subsubsection{Modeling of Non-Isolated Clusters}

Point sources in the vicinity of the extended source must be included in the
fit. We use local maxima in the combined wavelet scales 1--3 to create the
list of point sources. For the fitting, point sources are modeled as the PSF
calculated for a typical source spectrum as a function of off-axis angle.
Point source fluxes are free parameters, but their positions are fixed,
because they are accurately determined by the wavelet decomposition.  The
fitting procedure is analogous to that for isolated extended sources.

As was discussed above, some point sources are detected at scale 4, and
therefore we initially fit them as extended sources, i.e.\ by the
$\beta$-model with free core-radius and position. The best fit core radii
for such sources are small and consistent with zero, so they are not
included in the final catalog. However, these sources may interfere with the
determination of significance of source extent. Suppose that a faint point
source is located next to a bright cluster, and that the point source is
fitted by the $\beta$-model with free position. The best fit core-radius of
the point source component will be close to zero. To estimate the
significance of the cluster extent, we set the core-radius of the cluster
component to zero and refit all other parameters, including source
positions. In this case the best fit model will consist of the former
cluster component at the position of the point source and the former point
source component at the position of the cluster having non-zero core radius.
The net change of $-2\ln L$ will be zero and we will conclude that the
cluster component is not significantly extended. To overcome this
interference, we update source lists after the first fitting.  Those
extended sources which have best fit core radii $<5\arcsec$ are removed from
the list of extended sources and added to the list of point sources.
Parameters of the remaining extended sources are then refitted.

\subsubsection{Final Source Selection}

Next, we make the final selection of extended sources. 

1.~The main requirement is that the source must be real and significantly
extended. For this, we require that the formal significance of the source
existence must exceed $5\sigma$ and the significance of its extent must be
greater than $3.5\sigma$. 

2.~We find, however, that because of the non-linearity of the model, the
formal significance of the source extent is often overestimated for faint
sources on top of the very low background. To exclude these cases, we
required that the total source flux must exceed 25 photons.

3.~Some bright sources have a small but significant intrinsic extent. An
example is a bright Galactic star with a very soft spectrum. Its image is
slightly broader than the PSF for hard point sources because the PSF is
broader at low energies and the stars have a larger proportion of soft
photons. To exclude such cases, we required that the source core-radius must
be greater that 1/4 of the FWHM of the PSF. This requirement is met
automatically for faint clusters, because faint sources with small core
radii cannot be significantly extended, i.e.\ cannot satisfy condition (1).
This third criterion sets the lower limit of 6.25\arcsec\ for core-radii of
clusters in our catalog. Even at $z=1$, this angle corresponds to 50 kpc.

4.~Finally, one has to exclude sources associated with the target of
observation, as well as sources detected at large off-axis angles where PSF
degradation makes detection of the source extent uncertain. Our last
requirement was that the source is at least 2\arcmin\ from the target of the
observation and at off-axis angle smaller than 17.5\arcmin.

Sources satisfying criteria 1--4 comprise the final catalog.

\begin{figure*}[htb]
\mbox{}\hfill \includegraphics[width=3.5in]{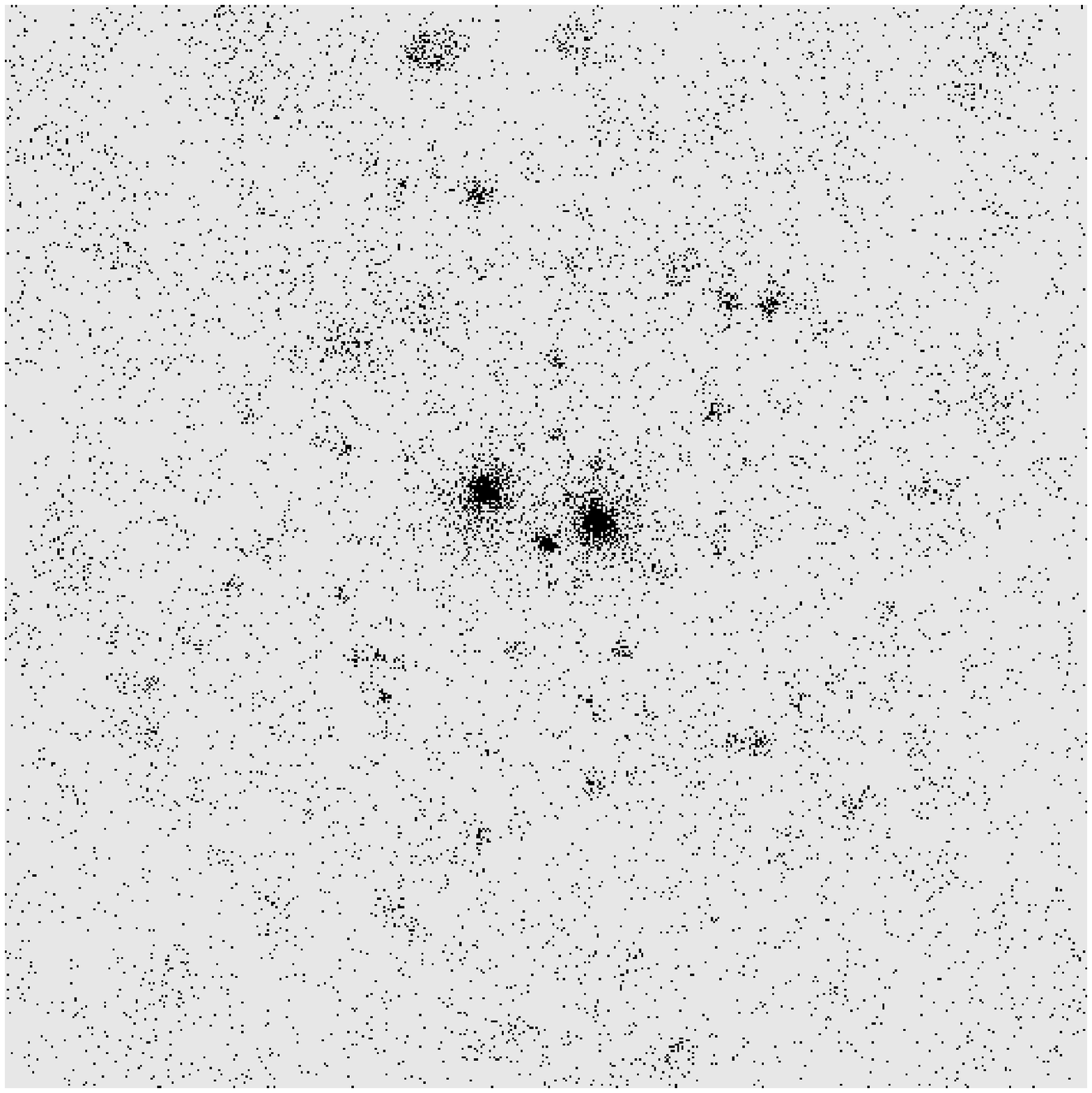} \hfill
\includegraphics[width=3.5in]{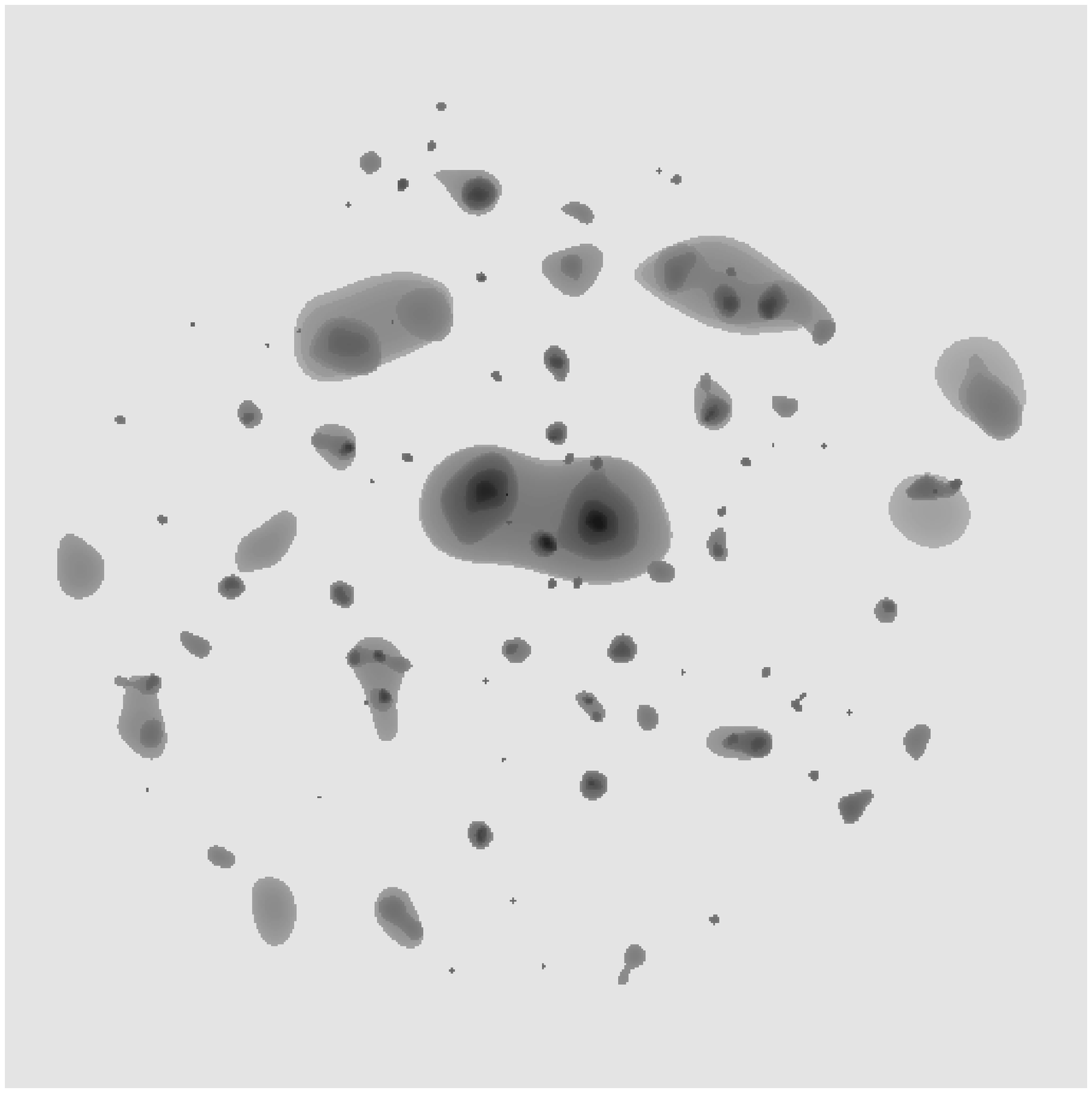} \hfill \mbox{}\par
\medskip
\mbox{}\hfill \includegraphics[width=3.5in]{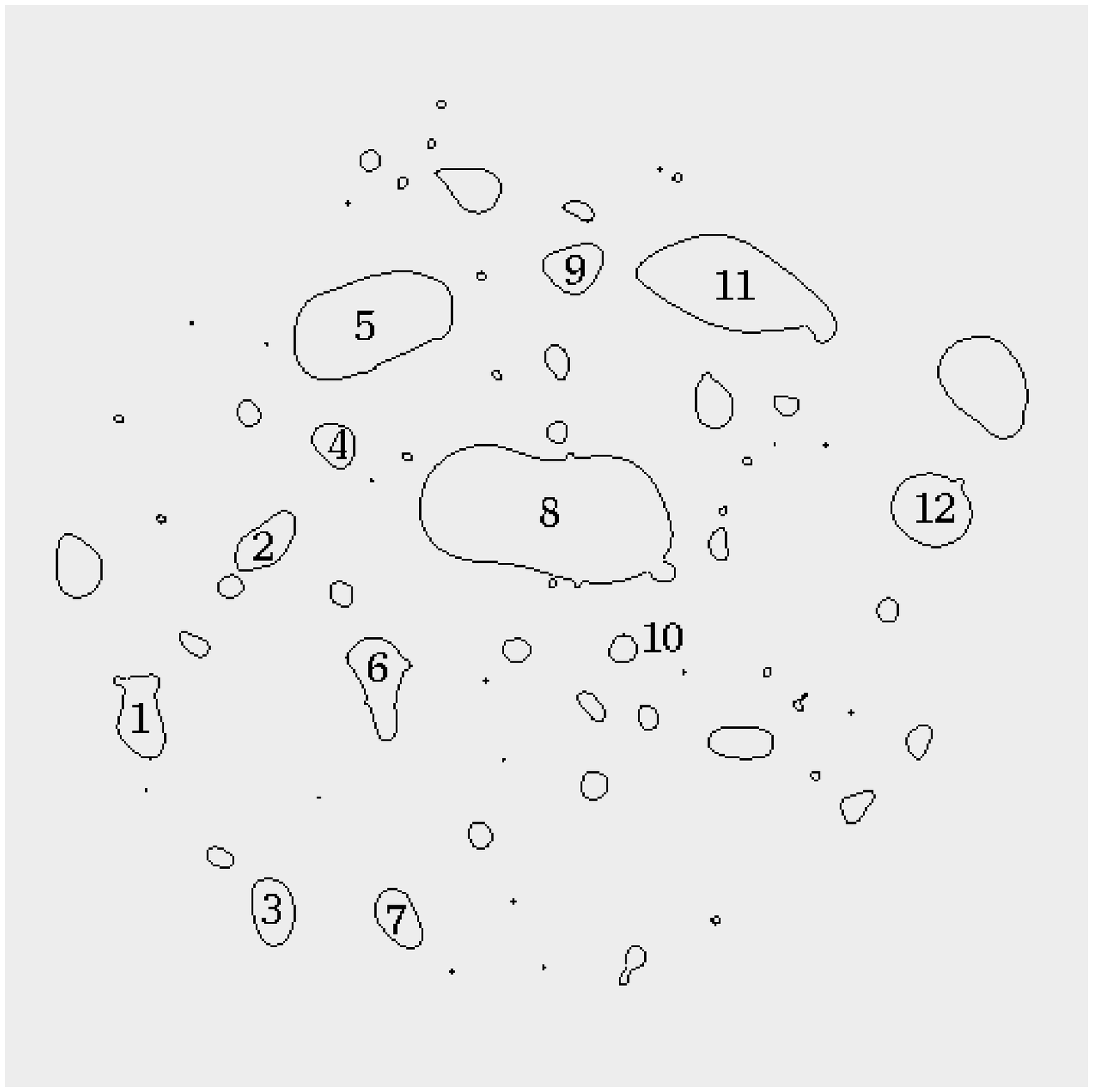} \hfill
\includegraphics[width=3.5in]{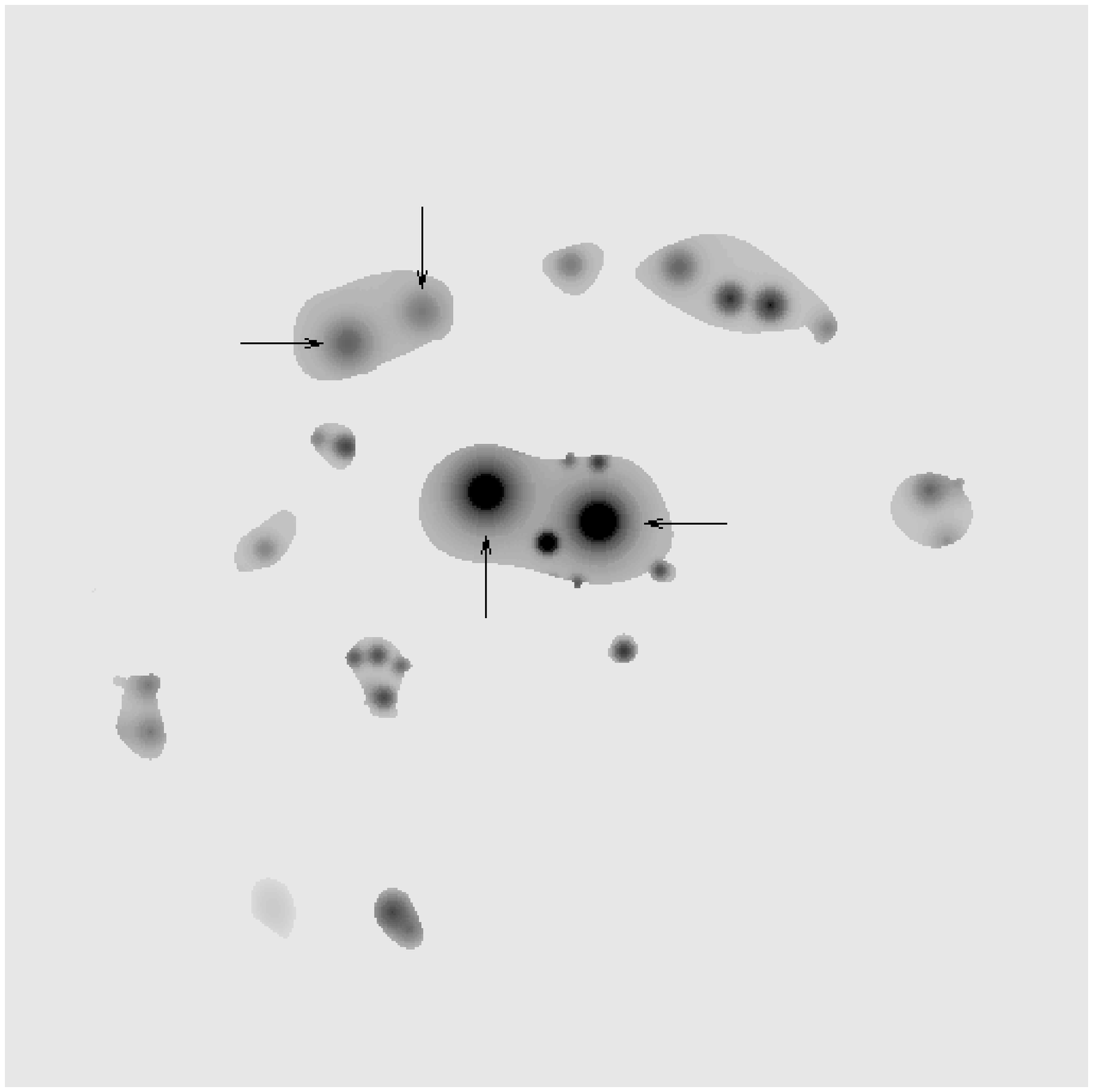} \hfill\mbox{}
\caption{\footnotesize 
Detection of extended sources in the 1701+6411 field. The wavelet
decomposition uses the photon image (\emph{a}) to detect significant
structures of different angular scale (\emph{b}). The wavelet image is split
into a number of connected domains (\emph{c}). The domains containing
candidate extended sources are numbered. The best fit image is shown in
panel \emph{d}. Extended sources which passed our final selection are
marked. All four sources were later confirmed as clusters by optical
observations.}
\label{fig:example}
\vskip -1.5ex
\end{figure*}

\subsection{A Real-Life Example}

To minimize computations, we fit the data only in those regions where the
sum of scales 1--6 is positive, i.e.\ where an excess over the background is
found by the wavelet decomposition.  To improve the computational efficiency
still further, the image is split into connected domains. Sources located
within the same domain are fit simultaneously.  The whole procedure of the
extended source detection is illustrated in Fig~\ref{fig:example}. The raw
photon image is shown in panel (\emph{a}).  The wavelet decomposition
detects 97 sources in this field.  The sum of scales 1--6 is shown in
Fig~\ref{fig:example}b. This image is split into connected domains
(Fig~\ref{fig:example}c). Domains which contain sources detected at scales
4, 5, or 6, are numbered. The best-fit model image in these domains is shown
in Fig~\ref{fig:example}d. Extended sources which passed the final
selection, are marked by arrows. All four of them are optically confirmed
clusters. Note that the number of candidate extended sources found by the
wavelet decomposition is more than 3 times the number of finally selected
clusters. Thus, the selection of candidate sources by the wavelet analysis
is rather lenient and does not miss real extended sources.

Using the detection procedure described in this section, we selected 239
significantly extended X-ray sources in 647 fields. In the following
sections we describe the measurement of their X-ray parameters, optical
observations, and present our final catalog.

\section{Measurement of Cluster X-ray Parameters}

For each detected cluster, we derive its position, radius, total X-ray flux,
and their uncertainties. All these quantities are derived from the best-fit
$\beta$-model, and their statistical errors are determined by Monte-Carlo
simulations. For this, we use the best-fit model image (which includes
clusters, point sources, and the background) as a template, simulate the
data using Poisson scatter, and refit the simulated data. The errors are
determined from the distribution of the best fit values in 100 simulations.
In this section, we discuss the measurement details and sources of
additional systematic errors of the cluster parameters.

\subsection{Positional Accuracy}\label{sec:positions}

Cluster position is measured as the best-fit centroid of the $\beta$-model.
In addition to the statistical uncertainty of the position, there is a
systematic uncertainty due to inaccuracy of the \ROSAT\/ aspect solution.
The aspect solution errors result in a systematic offset of all X-ray
sources in the field with respect to their optical counterparts. To correct
this error, we examined the positional correspondence of X-ray sources and
objects in the Digitized Sky Survey (DSS). If possible, targets of
observations or other prominent sources (galaxies or bright stars) were used
to find the precise coordinate correction. Coordinate shifts measured this
way have an uncertainty of 2\arcsec--5\arcsec, which is negligible compared
to the statistical error of cluster positions. If no optical counterparts of
X-ray sources were found in the DSS, we assigned a systematic position error
of 17\arcsec, the \emph{rms} value of shifts measured using targets of
observation. In some observations without a bright target, we found a
correlation between fainter X-ray and optical sources, and measured shifts
from this correlation. We regarded this shift measurement as less reliable
than that using targets, and assigned an intermediate systematic error of
10\arcsec\ to the cluster position in such fields.  The uncertain rotation
of the PSPC coordinate system results in a systematic error of $\sim
5\arcsec$ or less (Briel et al.\ 1996). We did not correct for the rotation,
but simply added $5\arcsec$ in quadrature to the offset uncertainty. The
final position error listed in Table~\ref{tab:catalog} is the sum of
systematic and statistical errors in quadrature.

\subsection{Core-Radius}\label{sec:radius}

Since it is impossible to fit the $\beta$-parameter using our data, we
measure core-radius for fixed $\beta=0.67$ and refer to this value as the
effective cluster radius $r_e$. Effective radius can be also defined as the
radius at which the surface brightness falls by a factor of $2^{3/2}$ and
hence is a physically meaningful combination of core-radius and $\beta$.
The $r_e$ measurement by fitting a $\beta=0.67$ model is accurate to $\pm
20\%$ within the observed range of $\beta$, $0.6<\beta<0.8$ (Jones \& Forman
1998).

We will now show that the radius measurement is relatively insensitive to
the presence of cooling flows which cause a surface brightness excess in the
central region of the cluster (e.g.\, Fabian 1994). Cooling flow clusters in
general cannot be fit by the $\beta$-model. However, in distant clusters,
the cenral excess is completely removed by the PSF blurring, and cooling
flows simply reduce the core-radius value. To study the possible influence
of the cooling flow on the derived effective radii, we use the \ROSAT\/ PSPC
image of Abell~2199, a nearby cluster with a moderate cooling flow of
$200\,M_\odot\,$yr$^{-1}$ (Edge et al.\ 1992). The $\beta$-model fit for all
radii yields $\beta=0.57$, $r_c=69\,$kpc. If the inner 200~kpc region is
excluded to remove the cooling flow contamination, the best-fit parameters
are $\beta=0.64$, $r_c=137\,$kpc, which corresponds to an effective radius
of $142\,$kpc. We then determine the radius value which we would measure if
A2199 were located at $z=0.4$. At this redshift, the FWHM of the PSF
corresponds to $\sim 200$~kpc.  We convolve the image with this ``PSF'' and
fit accounting for the smoothing and without exclusion of the center. The
best fit parameters for the smoothed data are $\beta=0.61$, $r_c=95\,$kpc.
Fixing $\beta=0.67$, as we do for the analysis of distant clusters, we
obtain $r_c=110$~kpc, only 22\% smaller than the true value obtained by
excluding the cooling flow.

\subsection{X-ray Flux}\label{sec:fluxcalc}

The surface brightness of most of detected extended sources significantly
exceeds the background only in a very limited area near the source center.
Therefore, the total source flux simply integrated in a wide aperture has
unacceptably large statistical uncertainty. To overcome this, the flux is
usually directly measured within some small aperture, and then extrapolated
to infinity using a reasonable model of the surface brightness profile
(Henry et al.\ 1992, Nichol et al.\ 1997, Scharf et al.\ 1997). Similarly to
this approach, we derived total fluxes from the normalization of the best
fit $\beta$-model. The most serious problem with the flux measurement using
such limited aperture photometry is the necessity to extrapolate the
observed flux to infinity. This extrapolation is a potential source of large
systematic errors because the surface brightness distribution at large radii
is unknown.  For example, consider the flux extrapolation from the inner 2.5
core radius region using $\beta$-models with different $\beta$.  This inner
region contains 49\% of the total flux if $\beta=0.6$, 64\% if $\beta=0.67$,
and 70\% if $\beta=0.7$.  Therefore, assuming $\beta=0.67$ one
underestimates the flux by $\sim 30\%$ if in fact $\beta=0.6$, the median
value in the Jones \& Forman (1998) sample. In addition, a trend of $\beta$
with cluster redshift or luminosity will introduce systematic changes within
the sample.  For example, Jones \& Forman find that lower luminosity
clusters have smaller $\beta$, which might result in underestimation of
their fluxes.

To address the issue of systematic flux errors in more detail, we have used
simulated realistic data (\S\ref{sec:simulations}) to estimate the effect of
the assumed value of $\beta$ on the cluster flux determination. Clusters
were fit as described in \S\ref{sec:ML}, but for three different values of
$\beta$, 0.6, $0.67$, and $0.7$. Dashed lines in Fig~\ref{fig:fluxbias} show
average ratios of the measured and input total flux as a function of the
true $\beta$, if the flux is measured as a normalization of the best-fit
model with $\beta$ fixed at 0.6 and 0.7.  In all cases significant biases
are present over the observed range of $\beta$ (Jones \& Forman 1998; shaded
region). We are interested in a flux measure which has the smallest
uncertainty for the whole range of $\beta$, not the one which yields an
unbiased flux estimate for some fixed value of $\beta$. The quantity
$(f_{0.6}+f_{0.7})/2$, where $f_{0.6}$ and $f_{0.7}$ are cluster fluxes
calculated assuming $\beta=0.6$ and $0.7$, respectively, is close to the
desired flux measure (solid line in Fig~\ref{fig:fluxbias}). It provides a
satisfactory flux estimate, accurate to $\pm10\%$ over the observed range of
$\beta$. We use this quantity to measure cluster fluxes throughout the rest
of this paper, and add the systematic error of $10\%$ to the statistical
uncertainty in the flux.

\begin{table*}
\vspace*{-1ex}
\tabcaption{\centerline{Comparison of flux measurements}\label{tab:comparison}}
\begin{center}
\renewcommand{\arraystretch}{1.2}
\footnotesize
\let\ph\phantom
\begin{tabular}{ccccccccc}
\hline
\hline
Cluster &  $z$ & Our survey &EMSS        & Nichol et al.& WARPS &
\multicolumn{3}{c}{Flux ratio$^{a}$} \\
\cline{7-9}
        &      & 0.5--2 keV  &0.3--3.5 keV& 0.3--3.5 keV & 0.5--2 keV &
EMSS & Nichol et al. & WARPS \\
\hline
MS 1201.5+2824 & 0.167 & 102.6       & 169.4       & 174.7       & 95.6 &
1.03  & 1.00 & 1.07 \\
MS 1208.7+3928 & 0.340 & \ph{1}26.6  & \ph{1}41.1  & \ph{1}42.7  & 29.3 &
1.12 &  1.08 & 0.91 \\
MS 1308.8+3244 & 0.245 & \ph{1}46.7  & \ph{1}69.3  & \ph{1}74.9  & 50.7 &
1.16 &  1.07 & 0.92 \\
MS 2255.7+2039 & 0.288 & \ph{1}50.5  & \ph{1}57.6  & \ph{1}73.9  & 51.9 &
1.53 &  1.19 & 0.97 \\
Average        &       &             &             &             &      &
1.21 &  1.09 & 0.97 \\
\hline
\end{tabular}
\end{center}
\footnotesize
$^a$ Ratios of fluxes measured in our survey and EMSS, Nichol et al., and
WARPS. To calculate these ratios, 0.3--3.5~keV fluxes were converted to the
0.5--2~keV energy band using the conversion coefficients from Jones et al.\
(1998).
\vskip -2.5ex
\end{table*}

Our sample includes four EMSS clusters (Henry et al.\ 1992), which were also
detected in the WARPS survey (Jones et al.\ 1998) and whose \ROSAT\/
observations were studied by Nichol et al.\ (1997). We use these clusters to
compare fluxes from all these surveys. Table~\ref{tab:comparison} shows
general agreement, within 10\%, between different \ROSAT\/ surveys,
especially between ours and WARPS. However, Henry et al.\ and, to a smaller
degree, Nichol et al.\ find fluxes which are systematically lower than those
from our survey and WARPS. Note that all
\ROSAT\/ surveys use essentially the same data, so the difference cannot be
explained by statistical fluctuations. Jones et al.\ have earlier performed
a similar comparison using a larger number of clusters. They also noted the
systematic difference of their fluxes compared to EMSS and Nichol et al.,
and explained this by the difference in flux measurement methods. All the
surveys derived fluxes by extrapolation from that measured within some
aperture using a $\beta$-model.  However, Henry et al.\ and Nichol et al.\
assumed fixed $\beta=0.67$ and $r_c=250$~kpc, while Jones et al.\ estimated
core-radii individually for each cluster, similar to our procedure.  Also,
our fluxes can be $\sim 5\%$ higher than those obtained for $\beta=0.67$,
because our measurements are optimized for the entire observed range of
$\beta$.  Cluster-to-cluster variations of $\beta$ probably explain $\sim
10\%$ non-systematic differences in flux for the same cluster in different
surveys. Jones et al.\ also compared their measurements with fluxes directly
integrated in a 4~Mpc aperture. They found that their fluxes exceed the
directly measured values by 10\%, with $\sim 60\%$ of that difference
explained by the cluster luminosity originating from outside 4~Mpc. Since
our measurements are $\sim 3\%$ lower than those of Jones et al., we
conclude that our fluxes are accurate within a few percent which is better
than the assigned systematic uncertainty.

\vskip -0.5ex
\centerline{\includegraphics[width=3.25in]{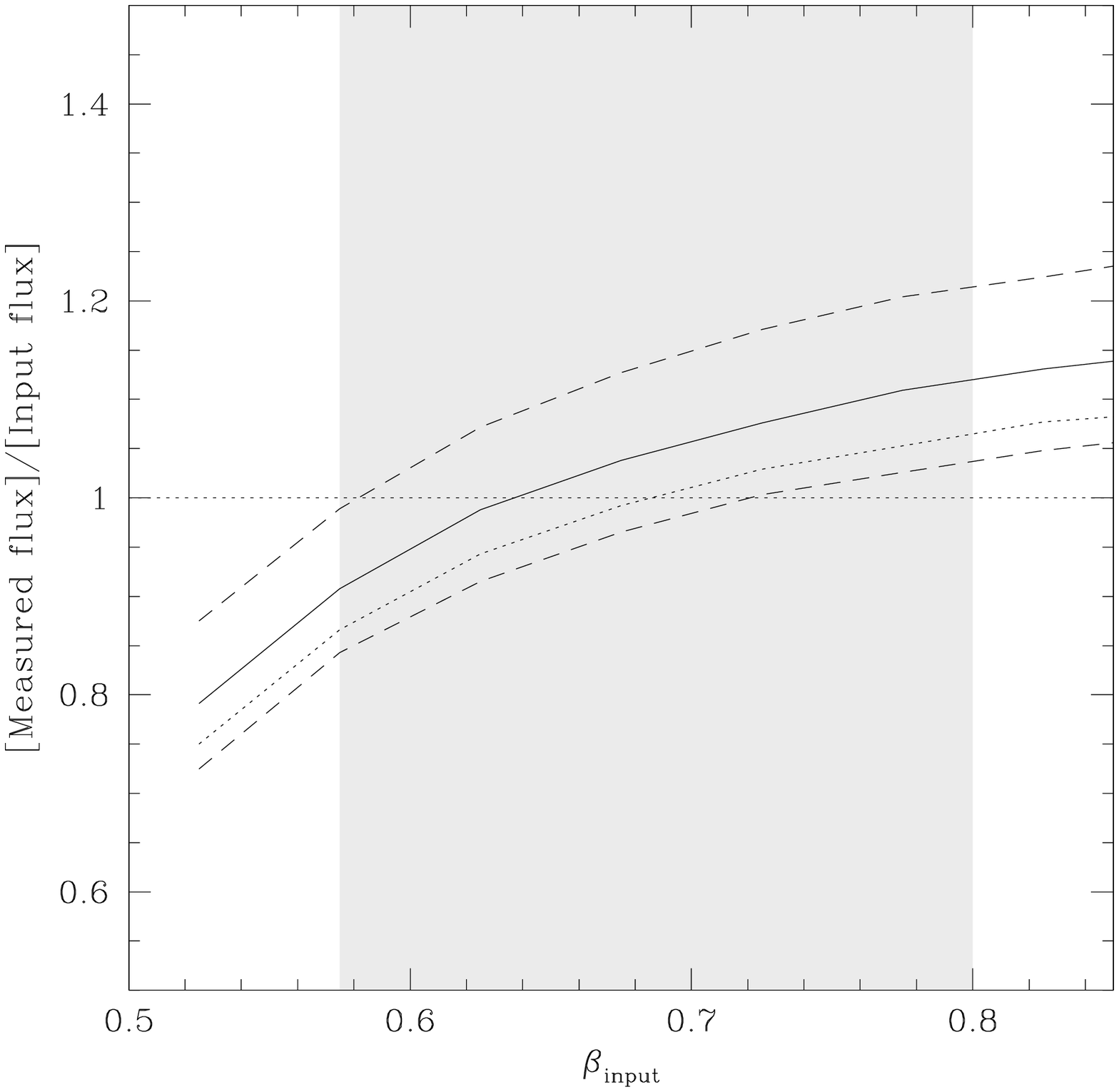}}
\vskip -2.5ex
\figcaption{Ratio of the measured and input cluster flux as a
function of the cluster $\beta$. Fluxes of simulated clusters were measured
by fitting $\beta$-models with $\beta$ fixed at 0.6 (upper dashed line), 0.7
(lower dashed line) and 0.67 (dotted line). Solid line corresponds to the
flux measure $(f_{0.6}+f_{0.7})/2$ used for our sample.
\label{fig:fluxbias}}

\section{Optical observations}

We are carrying on a program of optical photometric and spectroscopic
observations of our clusters. A complete discussion of optical observations
and data reduction will be presented in McNamara et al.\ (in preparation).
Below we discuss the optical results relevant to the X-ray catalog presented
here.

\subsection{Cluster Identification}

In some earlier works, optical identification of X-ray selected clusters
seeks a concentration of galaxies in redshift space, which requires a large
investment of telescope time. For our sources, the detected extended X-ray
emission is already a strong indication of cluster existence. Therefore, we
relaxed the optical identification criteria and required that either a
significant enhancement in the projected density of galaxies be found or
that an elliptical galaxy not included in the NGC catalog lie at the peak of
the X-ray emission.  While the galaxy concentration criterion is obvious,
the elliptical galaxy one is needed to identify poor clusters and groups
which fail to produce a significant excess of galaxies over the background.
It also helps to identify ``fossil groups'', in which galaxies have merged
into a cD (Ponman et al.\ 1994).  A potential problem with this second
criterion is that an active nucleus of an elliptical galaxy might be falsely
identified as a cluster.  However, a significant extent of X-ray emission in
all our sources makes this unlikely. Also, our spectroscopic observations of
such single-galaxy sources never showed emission lines characteristic of
AGNs.

\tabcaption{\centerline{Status of optical identifications}\label{tab:optidsum}}
\begin{center}
\renewcommand{\arraystretch}{1.2}
\footnotesize
\begin{tabular}{lr}
\hline
\hline
\multicolumn{2}{c}{Total sample} \\
Objects & 223 \\
Confirmed clusters & 200\\
False X-ray detections & 18 \\
No CCD imaging data & 5 \\ \\
\multicolumn{2}{c}{NED identifications} \\
Previously known clusters & 37\\
Previously known clusters with measured redshift & 29\\
NED AGN & 1\\
\\
\multicolumn{2}{c}{X-ray flux $>2\times10^{-13}\,$\ergs} \\
Objects & 82 \\
Confirmed clusters & 80\\
False detections & 1 \\
No data & 1 \\
\hline
\end{tabular}
\end{center}
\medskip

We obtained R, and in some cases I, V, and B band CCD images on the FLWO
1.2m, Danish 1.54m, and Las Campanas 1m telescopes. For brighter clusters,
we also used second generation Digitized Sky Survey (DSS-II) plates. Using
the DSS-II, it is possible to identify clusters at $z\lesssim0.45$. The
sensitivity of our CCD images is adequate to identify clusters to
$z=0.7-0.9$. If no cluster was visible in the CCD image, we considered this
object as a false detection (although it could be a very distant cluster).
These objects were retained in the sample for statistical completeness, but
marked in Table~\ref{tab:catalog}.

We also searched for possible optical counterparts in the NASA Extragalactic
Database (NED). The summary of NED identifications is given in
Table~\ref{tab:optidsum}. We obtained CCD photometry for some of the
catalogued clusters and tried to obtain spectroscopic data if redshifts were
not available. Fifteen extended sources were identified with isolated NGC
galaxies, and therefore removed from the cluster catalog. One object,
identified with an AGN, was considered as a false detection but was left in
the catalog for statistical completeness.

A summary of optical identifications of our cluster catalog is given in
Table~\ref{tab:optidsum}. In total, we confirmed 90\% of sources as clusters
in the total sample, while 8\% of sources are likely false detections. For
2\% of sources, no optical counterpart was present in the DSS-II and no CCD
images were yet obtained. In the X-ray bright subsample,
$f>2\times10^{-13}\,$\ergs, we optically confirmed 98\% of sources as
clusters; one object in this subsample is a false detection and for the
remaining one, optical images are saturated by neighboring Arcturus.  These
high success rates demonstrate the high quality of our X-ray selection.

\begin{figure*}
\vspace*{-3ex}
\mbox{}\hfill\includegraphics[width=3.25in]{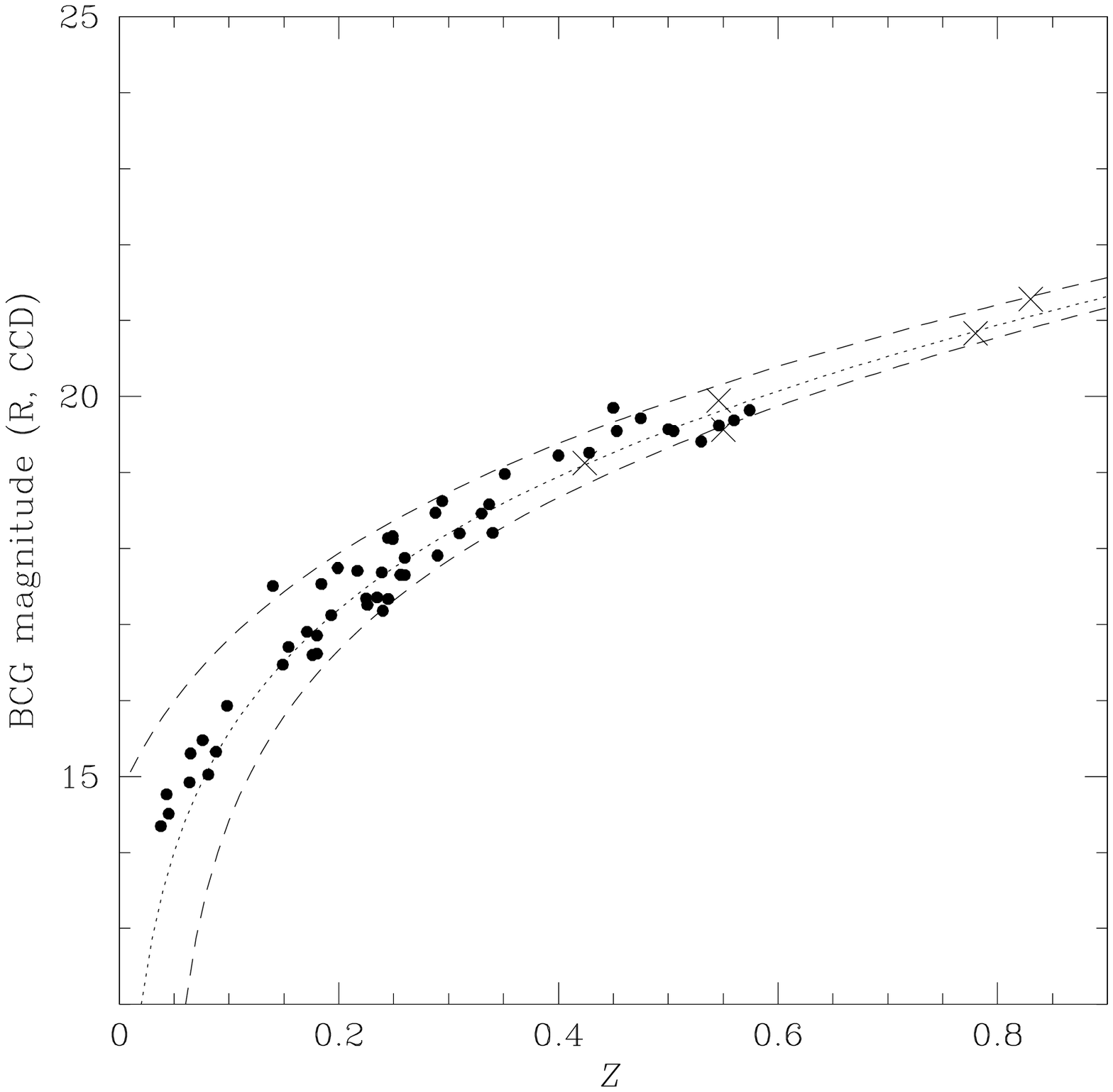}\hfill
	     \includegraphics[width=3.25in]{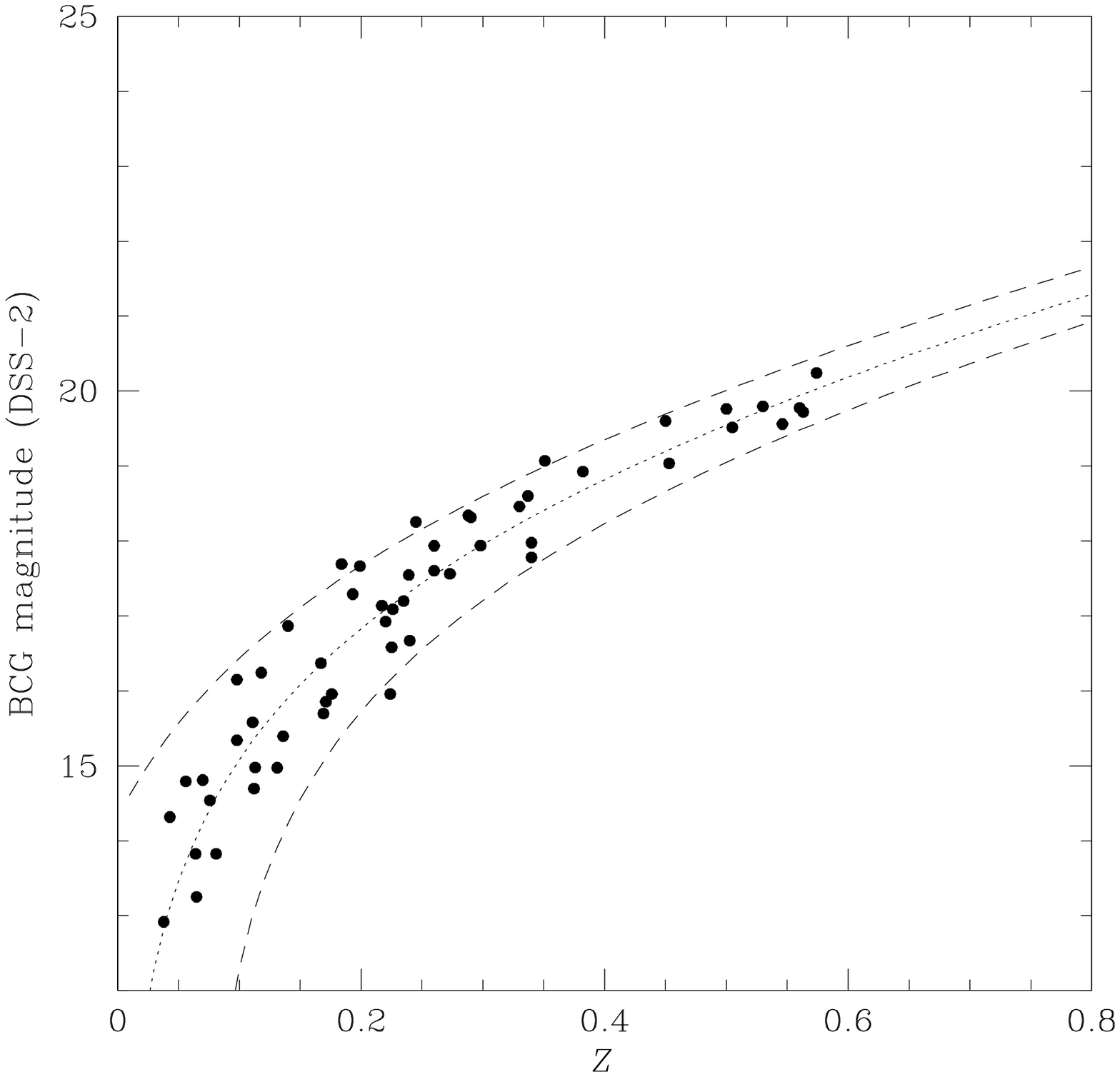}\hfill\mbox{}
\vskip-4ex
\figcaption{(\emph{a}) X-ray luminosity corrected BCG magnitudes vs.\
redshift. The dotted line shows the analytical fit (see text). The estimated
redshift uncertainty of $\Delta z = ^{+0.04}_{-0.07}$ is shown by dashed
lines. Crosses mark five high-redshift EMSS clusters. These clusters were
not used in the fit.\label{fig:magz:ccd} (\emph{b}) Same as (\emph{a}) but
magnitudes were measured using DSS-II. The dotted line shows the best fit
relation, and dashed lines correspond to $\Delta z =\pm 0.07$.
\label{fig:magz:dss}}

\vskip-2.5ex
\end{figure*}

\subsection{Spectroscopic and Photometric Redshifts}\label{sec:photz}

We observed an incomplete subsample of clusters spectroscopically on the
MMT, ESO 3.6m, and Danish 1.54m telescope. In most cases, we identified
several obvious cluster galaxies in the CCD images and then obtained a
long-slit spectrum, usually for 2--3 galaxies per cluster. The slit always
included the brightest cluster galaxy. For 10 clusters observed at the ESO
3.6m telscope, we obtained multi-object spectra, 10--15 galaxies per
cluster.  Altogether, we measured 47 redshifts ranging from $z=0.040$ to
$z=0.574$.  Further details of spectroscopic observations will be presented
in McNamara et al.\ (1998, in preparation).

For those clusters without spectroscopic data, we estimated redshifts from
the magnitude of the brightest cluster galaxy (BCG). The BCG was selected as
the brightest galaxy either within the error circle of the cluster X-ray
position or the one in the center of the galaxy concentration; both criteria
were met simultaneously in most cases. Although the BCG selection was
somewhat subjective, the tightness of the magnitude vs.\ redshift relation
obtained for $\sim 1/4$ of the total sample confirms our procedures. For
nearby clusters, the scatter in the absolute magnitude of BCGs is small,
$\sigma_M\approx0.2$ (Sandage 1972), which corresponds to $\approx 10\%$
distance error.  Our results show that the scatter is small at higher
redhifts as well. The magnitude vs.\ redshift relation is calibrated within
our sample, and photometric redshifts are estimated using the CCD images
obtained under photometric conditions or DSS-II plates.

The CCD galaxy photometry was performed in the R band. The BCG magnitudes
were measured within a fixed 4\arcsec aperture. Such an aperture was chosen
to make the measurement relatively insensitive to poor seeing, which was
$\sim 2\arcsec$ in some cases, and encompass $\sim 50\%$ of light in
high-redshift galaxies. The fixed angular aperture corresponds to the metric
aperture increasing with redshift, from 10~kpc at $z=0.1$ to 29~kpc at
$z=0.5$. The increase of the metric aperture is a monotonic function of
redshift, the same for all clusters, and thus does not prevent us from using
the $m-z$ relation for photometric redshift estimates.  We did not make
K-corrections of BCG magnitudes because this is also a monotonic systematic
function of redshift.  Measured magnitudes were corrected for Galactic
extinction using Burstein \& Heiles (1982) maps.

There is a correlation between the BCG magnitude and the cluster X-ray
luminosity (Hudson \& Ebeling 1997), which increases the scatter in the
$m-z$ relation. Within our sample, the absolute BCG magnitude changes
approximately as $-0.5\log L_x$, in good agreement with the Hudson \&
Ebeling results. Below we use the corresponding correction, $m^\prime = m +
0.5\log(L_x/10^{44}\,\mbox{erg s$^{-1}$})$ to compensate for this effect.

The X-ray luminosity-corrected BCG magnitude is plotted vs. cluster redshift
in Fig~\ref{fig:magz:ccd}. This dependence can be well fit by a cosmological
dimming law $m^\prime = m_0 + 5\log z - 1.086(q^\prime-1)z$ with best-fit
parameters $m_0=20.45\mag$ and $q^\prime=-0.121$.  In this equation,
$q^\prime$ provides a useful parametrization but does not have the meaning
of the cosmological deceleration parameter, because magnitudes were not
K-corrected and a varying metric aperture was used. The best fit relation is
shown by the dotted line in Fig.~\ref{fig:magz:ccd}. Photometric redshifts
were estimated from the analytical fit using the following iterative
procedure. We estimated redshift from the uncorrected BCG magnitude. Using
the estimated redshift, we calculated the X-ray luminosity, corrected the
BCG magnitude as described above, and re-estimated $z$. The process was
repeated until the estimated redshift converged. We checked this procedure
by estimating photometric redshifts of clusters with measured redshifts.
This comparison has shown that the photometric estimate is unbiased and has
an uncertainty of $\Delta z = ^{+0.04}_{-0.07}$.

We also observed five high-$z$ EMSS clusters (0302+1658, 0451.6--0305,
0015.9+1609, 1137.5+6625, and 1054.5--0321) to check the $m-z$ relation at
high redshift using an external X-ray selected sample.  These clusters are
plotted by crosses in Fig.~\ref{fig:magz:ccd}. They follow the relation
defined by our sample very well. In addition, these five EMSS clusters are
very X-ray luminous; their accordance with the $m-z$ relation confirms the
validity of the X-ray luminosity correction we apply to BCG magnitudes.

For 13 clusters without photometric CCD data, redshifts were estimated using
the Second Digitized Sky Survey plates. Photometric calibration of was
performed using our CCD images, and will be described in McNamara et al.\
(1998, in preparation). BCG magnitudes were measured in a fixed angular
aperture of 5\arcsec. No K-correction was applied. The X-ray luminosity
corrected DSS-II magnitudes are plotted vs.\ redshift in
Fig.~\ref{fig:magz:dss}. The $m-z$ relation can be fit by the relation $m =
m_0 + 5\log z - 1.086(q^\prime-1)z$ with best fit parameters $m_0=19.84$,
$q^\prime=-1.23$ photometric redshifts were estimated using a procedure
analogous to that for the CCD data. The comparison of the estimated and
measured redshifts yields the accuracy of the photometric estimate of
$\Delta z\approx \pm 0.07$.

\section{The Catalog}

Our cluster catalog is presented in Table~\ref{tab:catalog}. The object
number is given in column~1. The coordinates (J2000.0) of the X-ray centroid
are listed in columns 2 and 3. The total unabsorbed X-ray flux in the
0.5--2~keV energy band (observer frame) in units of $10^{-14}\,$\ergs\ and
it uncertainty are listed in columns 4 and 5. Angular core-radius and its
uncertainty are given in columns 6 and 7. Column~8 contains spectroscopic or
photometric redshifts.  The 90\% confidence interval of the photometric
redshift is given in column~9. Thirteen clusters for which the DSS was used
for photometric redshift are marked by superscript in column~9. If redshift
is spectroscopic, no error interval is given. Three clusters show clear
concentrations of galaxies near the X-ray position, but the choice of BCG is
uncertain because of the large cluster angular size. We do not list
photometric redshifts for these clusters and mark them by ``U'' in the Notes
column. Columns~10 lists 90\% X-ray position error circle.  Column~11
contains notes for individual clusters. In this column, we list the optical
identifications from the literature. We also mark likely false detections by
``F''.

Table~\ref{tab:lofields} shows coordinates and exposures for the 647
analyzed \ROSAT\/ pointings. For a quick estimate of sensitivity in each
field, one can use the listed exposure time and Fig~\ref{fig:limflux}.  In
this figure, we show the limiting flux, at which clusters are detected with
a probability of $90\%$ for off-axis angles between 2\arcmin\ and
17.5\arcmin.

\section{Monte-Carlo Simulations of Cluster Detection}\label{sec:simulations}

For a statistical analysis of our cluster catalog, the detection efficiency
as a function of flux and extent, and measurement uncertainty is required.
To derive these functions, we used extensive Monte-Carlo simulations
described in this section.

\subsection{Correcting for Selection Effects}

The most direct way to compare theoretical models with our cluster catalog
is to predict the number of clusters within some interval of measured fluxes
and radii (and redshift) and then compare the prediction with the number of
detected clusters in this interval. To predict the number of detected
clusters, one needs to know the detection probability as a function of real
cluster flux, $f$, and radius, $r_c$, and the distribution of measured
values, $f_m$ and $r_{c,m}$, also as a function of $f$ and $r_c$. Using a
theoretical model, one calculates the number of real clusters as a function
of flux and radius, then multiplies this number by the detection
probability, and then convolves it with the measurement scatter.  

Since the detection algorithm for extended sources is rather complicated,
the only method of deriving appropriate corrections is through Monte-Carlo
simulations.

\medskip
\centerline{\includegraphics[width=3.25in]{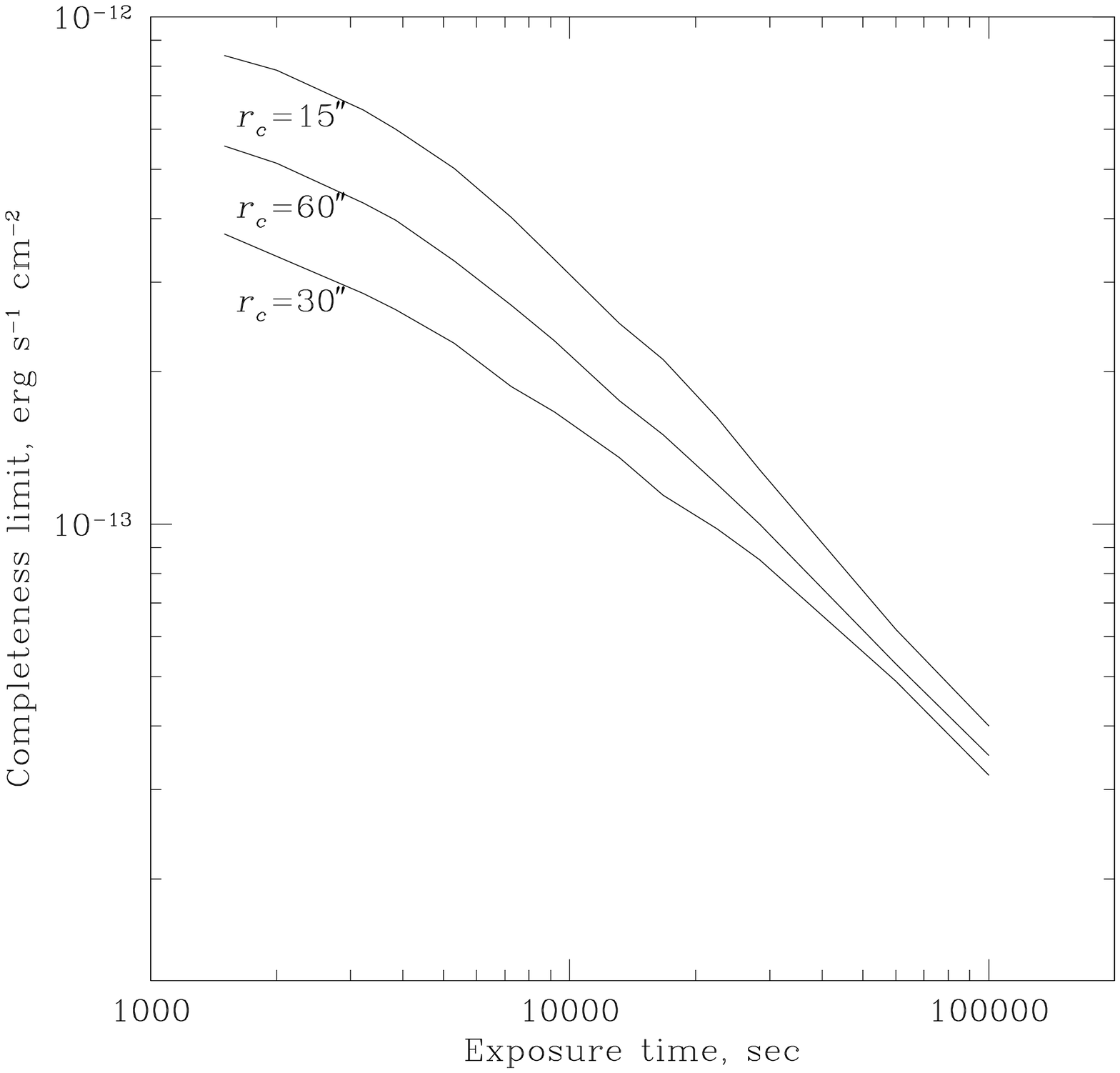}}
\vskip -3.5ex
\figcaption{Approximate limiting flux, at which the cluster detection
probability is 90\% in the range of off-axis angles $2\arcmin-17.5\arcmin$,
plotted vs.\ exposure time. Limiting fluxes for three values of cluster
core-radius, $r_c=15\arcsec$, 30\arcsec, and 60\arcsec, are shown.
Sensitivity is best for $r_c\approx 30\arcsec$ and declines for smaller and
larger clusters.
\label{fig:limflux}}

\subsection{What Affects the Cluster Detection?}

In this section, we discuss the effects that influence the cluster detection
process, and therefore should be included in Monte-Carlo simulations.

The first obvious effect is the degradation of the \ROSAT\/ angular
resolution at large off-axis angles. Because of this degradation, a cluster
with $r_c=20\arcsec$ is well-resolved on-axis where the FWHM of the PSF is
25\arcsec, but the same cluster is indistinguishable from a point source if
located at an off-axis angle of 17\arcmin\ where the PSF is 57\arcsec\
(FWHM).

Point sources, which may lie in the vicinity of a cluster, reduce the
efficiency of cluster detection and increase the measurement errors.
Therefore, the simulations should include realistic spatial and flux
distributions of point sources.

In addition, exposure time, Galactic absorption, and the average background
level vary strongly among the analyzed \ROSAT\/ fields, and so does the
probability to detect a cluster of given flux. Also, the background has to
be modeled individually for each field, and cannot be assumed known in
simulations.

To model all these effects, we simulate realistic \ROSAT\/ images containing
point sources, insert clusters with known input parameters at random
positions into the simulated images, and analyze these images identically to
the real data. The selection functions are then derived from comparison of
the numbers and parameters of input and detected clusters.

\subsection{Simulating ROSAT Images without Clusters}\label{sec:sim:point}

We begin with point sources which are the major contributor to the X-ray
background in the \ROSAT\/ band. To simulate source fluxes, we use the $\log
N - \log S$ relation measured in the flux range of
$1.2\times10^{-15}-10^{-12}\,$\ergs\ (Vikhlinin et al.\ 1995). Fluxes are
simulated using the extrapolation of $\log N - \log S$ in the range from
$10^{-11}$ to $2.5\times10^{-17}\,$\ergs, where the integral emission of
point sources saturates the X-ray background.  Source positions are
simulated either randomly or with a non-zero angular correlation function
using a two-dimensional version of the Soneira \& Peebles (1978) algorithm.
After the source position is determined, we convert the flux to the number
of detected photons using the exposure time at the source position, and the
counts-to-flux conversion appropriate to a power law spectrum with
$\Gamma=2$ and the actual Galactic absorption in the simulated field.  The
number of detected source photons is drawn from a Poisson distribution.  The
photon positions are simulated around the source position according to the
PSF as a function of off-axis angle. Finally, we add a flat Poisson
background (corrected for the exposure variations across the field) until
the average background levels are equal in the simulated image and the
corresponding real observation. This flat uniform component corresponds to
truly diffuse backgrounds, such as foreground Galactic emission, scattered
solar X-rays, and the particle background.

The images simulated according to the described procedure correctly
reproduce fluxes and the spatial distribution of point sources, the average
background level, and background fluctuations caused by undetected point
sources and their possible angular correlation.

\subsection{Simulations of clusters}

The next step is to put a cluster of a given flux and angular size at a
random position in the image. An elliptical $\beta$-model
\begin{equation}\label{eq:ellbeta}
I(x,y) = I_0\;\left(1+x^2/a_x^2+y^2/a_y^2\right)^{-3\beta+1/2},
\end{equation}
was used for cluster brightness. Cluster $\beta$ parameters and axial ratios
were randomly selected from the distribution observed in nearby clusters
(Jones \& Forman 1998, Mohr et al.\ 1995). To include the influence of edge
effects arising because detected clusters must lie between off-axis angles
of 2\arcmin--17.5\arcmin, cluster positions were simulated in the inner
18.5\arcmin\ circle of the field of view. Cluster flux was converted to the
number of detected photons using the local exposure and the counts-to-flux
coefficient corresponding to a $T=5$~keV plasma spectrum and the Galactic
absorption for the field. The cluster model was convolved with the PSF
calculated for the given off-axis angle. Photons were simulated using a
Poisson distribution around the model and added to the image.

Reducing simulated images identically to the real data, we derive the
detection efficiency as a function of flux and effective radius. Effective
radius is defined as the radius at which the radially-averaged surface
brightness drops by a factor of $2^{3/2}$ (\S\ref{sec:radius}). The
effective radius can be calculated from parameters in eq.~\ref{eq:ellbeta}
as
\begin{equation}
r_e=\sqrt{a_xa_y\left(2^{1.5/(3\beta-0.5)}-1\right)}.
\end{equation}
In simulations, we verified that the detection probability indeed has very
little dependence on the $\beta$-parameter and axial ratio and is determined
by $r_e$ only.

\subsection{Simulation Runs}

Simulated images were reduced identically to the real data, i.e.\ we modeled
the background (\S\ref{sec:bg}), detected candidate extended sources by the
wavelet decomposition (\S\ref{sec:wd}), fitted these candidate sources and
applied our final selection criteria (\S\ref{sec:ML}), and recorded
parameters of input and detected clusters. Each of the 647 \ROSAT\/ fields
was simulated 650 times. Radii and fluxes of input clusters were randomly
distributed in the 5\arcsec--300\arcsec,
$10^{-14}$--$3\times10^{-12}\,$\ergs\ range.

To derive the distribution of false detections, we performed a separate set
of simulations without putting clusters into simulated images.  In this set
of simulations, each field was simulated 50 times.

Simulations were performed with point sources distributed either randomly or
with the angular correlation function measured by Vikhlinin \& Forman (1995)
for faint \ROSAT\/ sources. The spatial correlation of point source
significantly increases the number of false detections (by a factor of 1.5),
but has little or no effect on the detection probability of real clusters.

The simulation results were used to measure the cluster selection functions
necessary for a statistical study of our catalog. These data are available
in electronic publication on AAS CDROM and through the WWW page
\mbox{http://hea--www.harvard.edu/x--ray--clusters/}.

\mbox{}\hfill\includegraphics[width=3.25in]{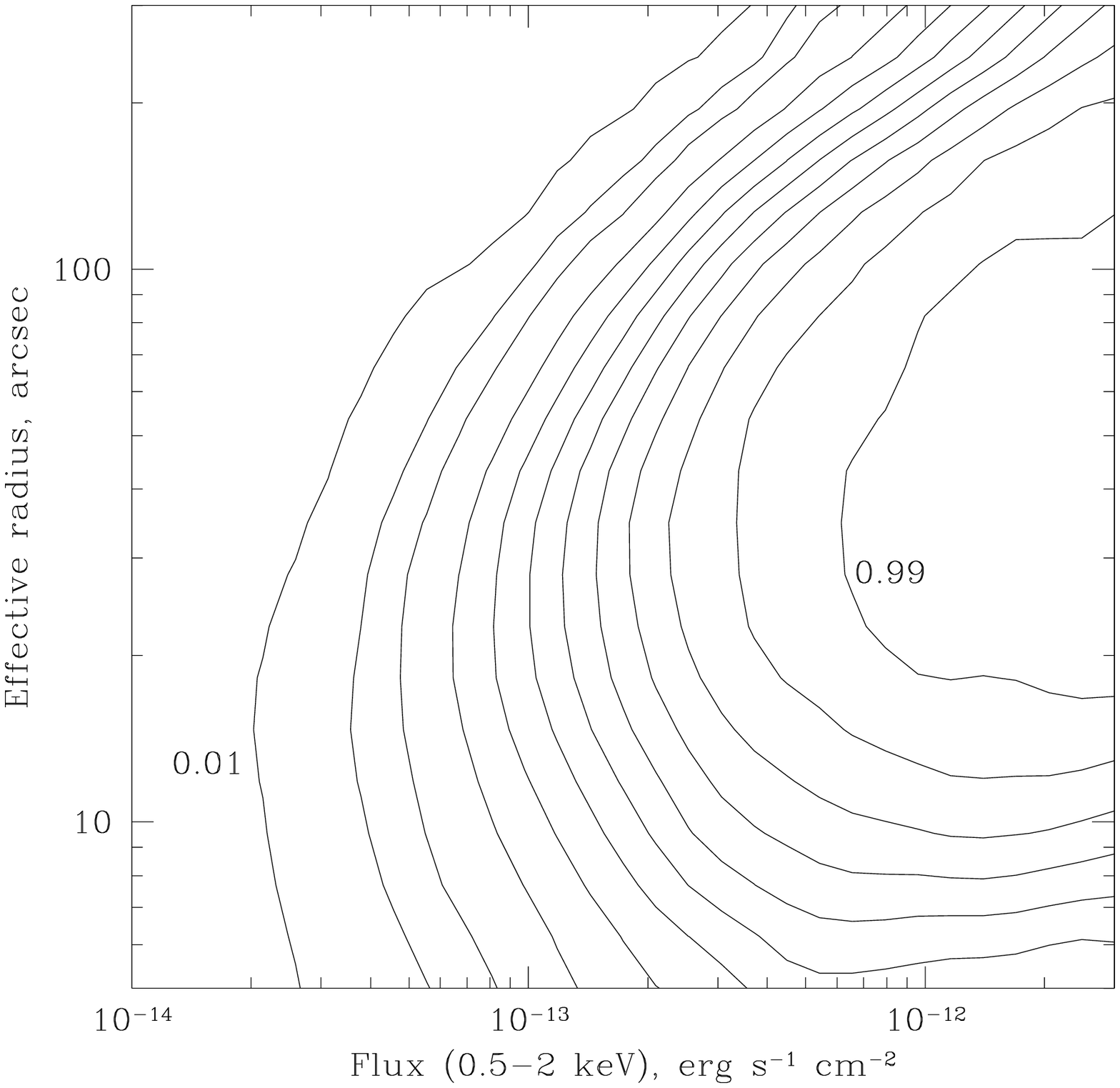}\hfill\mbox{}
\vskip-2.5ex
\figcaption{Probability of cluster detection as a function of flux
and core-radius. Contours correspond to the detection probabilities of 1, 5,
10, 20, 30, 40, 50, 60, 70, 80, 90, and 99\%. \label{fig:effarea2d}}

\subsection{Results of Simulations: Detection Probability}\label{sec:detprob}

The probability that a cluster with unabsorbed flux $f$ and radius $r_e$,
whose position falls within 18.5\arcmin\ of the center of one of the
analyzed \ROSAT\/ fields will be detected, is shown in
Fig.~\ref{fig:effarea2d}. This probability is normalized to the geometric
area of the annulus in which detected clusters may be located
(2\arcmin--17.5\arcmin). At a given flux, the detection probability is the
highest for clusters with radii of $\sim 30\arcsec$. It gradually decreases
for clusters with large radius, because their flux is distributed over the
larger area, thus decreasing their statistical significance. The detection
probability also decreases for compact clusters, because they become
unresolved at large off-axis angles. This effect is important for clusters
with angular core radii of $\lesssim 15\arcsec$. Even at $z=1$ this radius
corresponds to 130~kpc, which is two times smaller than the core-radius of a
typical rich cluster (250~kpc, Jones \& Forman 1998).  Therefore, cluster
detection efficiency is limited mainly by the low number of photons, not by
the resolution of the \ROSAT\/ PSPC.

The detection probability changes by less than 10\% for clusters with axial
ratios $<0.7$ compared to azimuthally symmetric clusters.  This is caused by
significant PSF smearing, which reduces the apparent ellipticity of distant
clusters.  Similarly, we have found no significant dependence of the
detection probability on the value of the $\beta$-parameter.

\begin{figure*}[htb]
\vspace*{-3ex}
\mbox{}\hfill\includegraphics[width=3.25in]{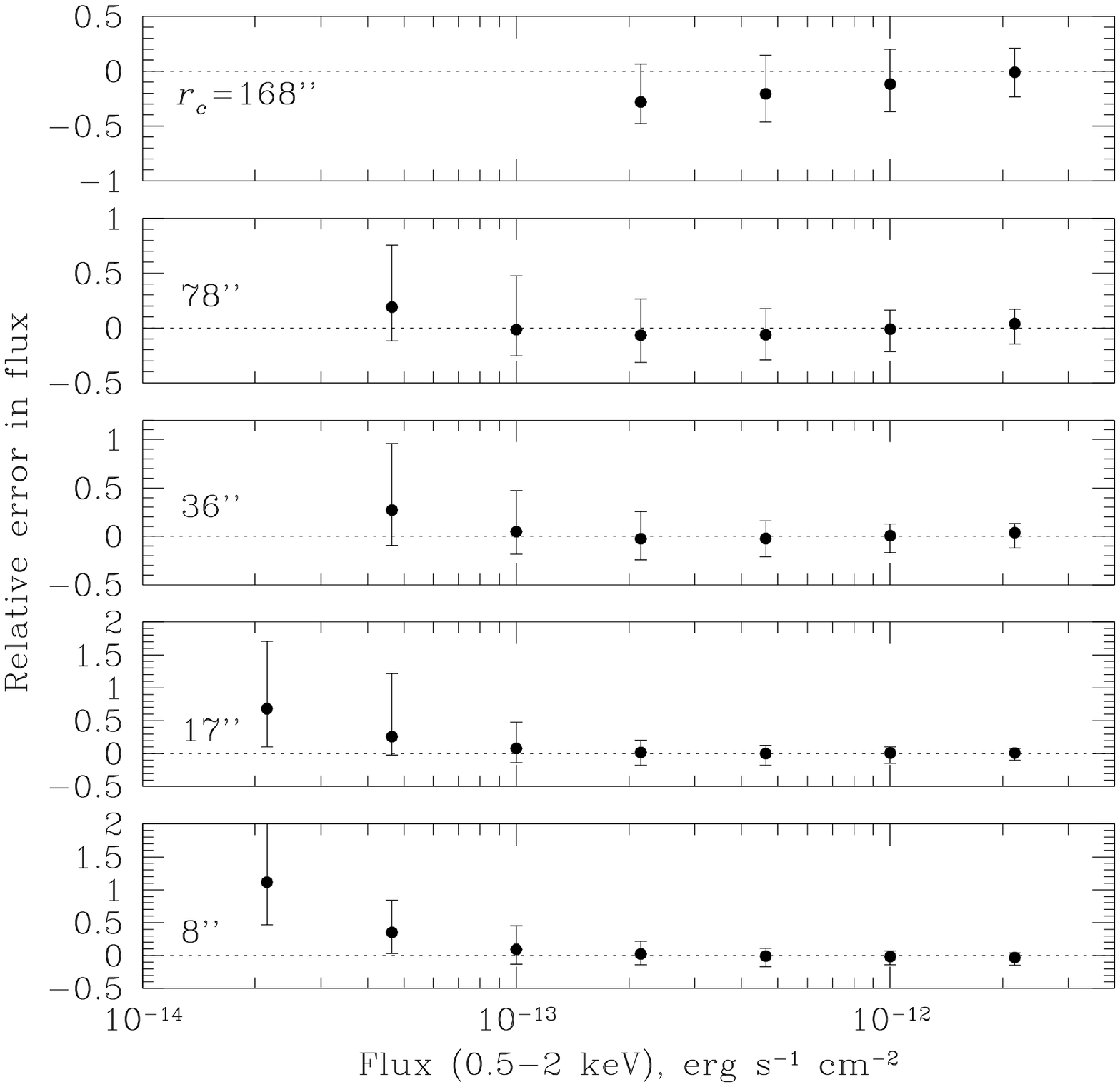}\hfill
	     \includegraphics[width=3.25in]{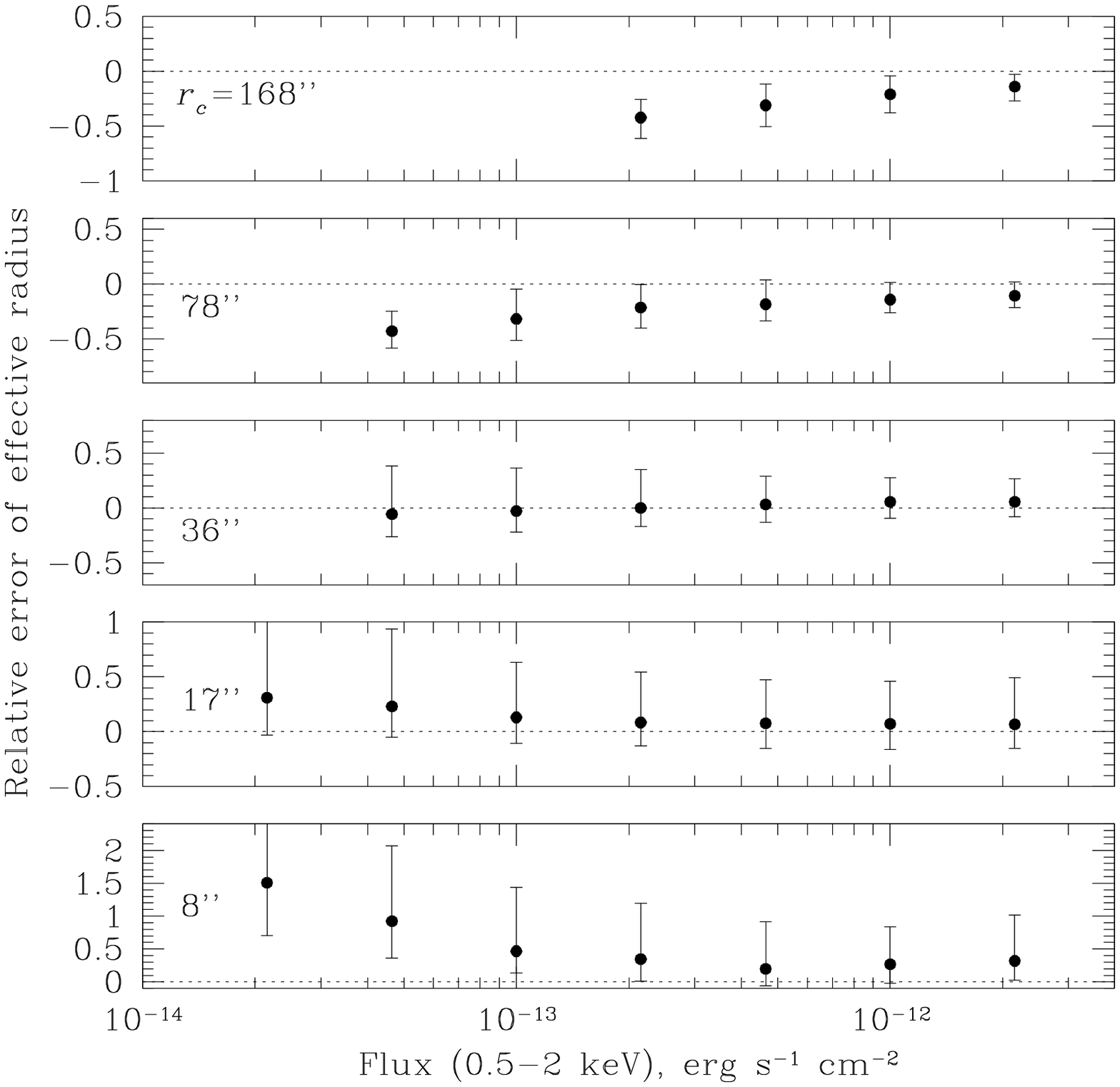}\hfill\mbox{}
\vskip -2.5ex
\caption{\footnotesize Bias and scatter of flux and radius measurements.
Points show the average relative deviation of the measured quantity. Error
bars show the relative scatter of the measured quantity, not errors of bias.}
\label{fig:biasgrid}
\vskip -1.5ex
\end{figure*}

\begin{figure*}[htb]
\vspace*{-2.5ex}
\mbox{}\hfill\includegraphics[width=3.25in]{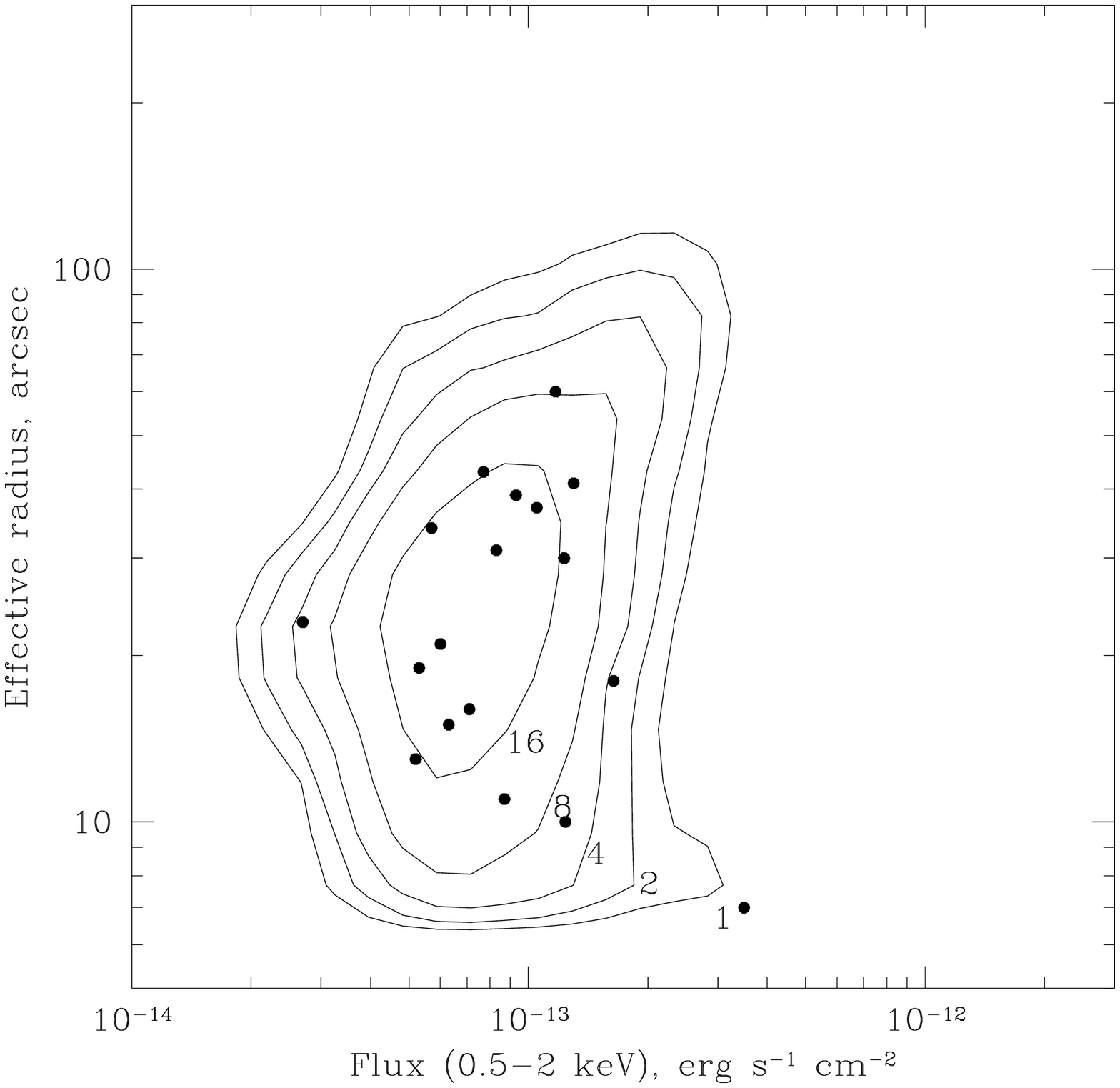}\hfill
	     \includegraphics[width=3.25in]{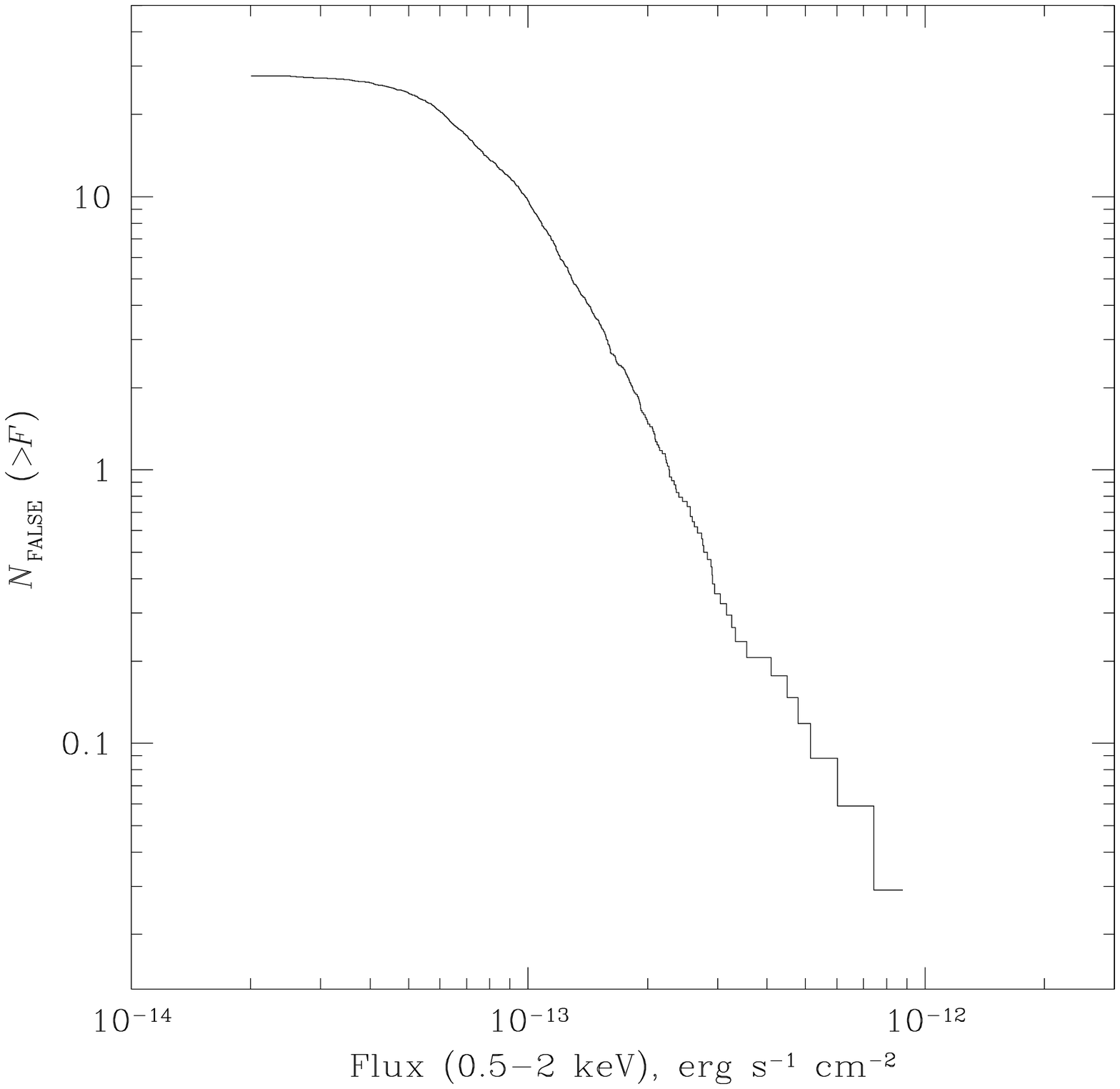}\hfill\mbox{}
\vskip -2.5ex
\caption{\footnotesize Distribution of false sources as a function of
measured flux and radius (left), and a cumulative distribution as a function
of flux (right). In the left panel, contours represent the levels of equal
density of false source distribution. Contour labels show the number of
false sources outside the contour. Points represent parameters of the likely
false detections in the data, i.e.\ those X-ray sources which have no
cluster counterparts in deep CCD images.}
\label{fig:falsedistr}
\vskip -1.5ex
\end{figure*}

\subsection{Results of Simulations: Measurement Scatter and Bias}\label{sec:measscat}

In this section we consider the distribution of measured flux and radius of
detected clusters, as a function of input flux and radius. This distribution
is derived for clusters detected in \emph{any}\/ field and at \emph{any}\/
off-axis angle, and is different from the uncertainties listed in
Table~\ref{tab:catalog}, which are determined only by the photon counting
statistics. Fig.~\ref{fig:biasgrid} shows the distributions derived for
several values of input cluster flux and radius. The points in this figure
represent the mean relative deviation of the observed parameter, while the
error bars show the mean relative scatter (both positive and negative).
Generally, the flux measurement is unbiased and has a small relative scatter
of $\sim 20\%$. At low fluxes, where the detection probability decreases,
the measured fluxes tend to be overestimated. This bias is naturally present
whenever a flux measurement is performed near the detection threshold, and
is not related to the particular detection algorithm. For example, a source
with a true flux exactly equal to the detection threshold will be detected
with 50\% probability, and in all these cases the measured flux exceeds the
true flux. Averaged over detections, the measured flux exceeds the true
value. This flux bias should be accounted for in deriving the luminosity
functions and the $\log N -
\log S$ relation. On the other hand, for clusters with very large radii, the
flux is underestimated, because the background is overestimated near broad
clusters.  This effect is important only for clusters at low redshifts which
have large angular core radii.

The radii of very compact clusters are strongly overestimated on average,
because such clusters can be detected as extended sources only if their
measured radius is a positive fluctuation with respect to the true value,
similar to the flux bias above.  The measured radii of very broad clusters
are underestimated because of the oversubtraction of the background.  The
sizes of distant clusters mostly fall in the range of $15\arcsec-1\arcmin$,
where our radius measurements are unbiased.  For example, a radius of
250~kpc corresponds to 45\arcsec{} at $z=0.3$, 35\arcsec{} at $z=0.5$, and
29\arcsec\ at $z=1$. At $z<0.2$, 250~kpc corresponds to large angular radii
and therefore measured sizes of large low-redshift clusters are
underestimated.

\subsection{Results of Simulation: False Detections}\label{sec:false}

Because of the finite angular resolution of the \ROSAT\/ PSPC, closely
located point sources can be falsely classified as a single extended source.
Optical identification is the most direct way of finding such false
detections. However, optical observations alone, with no estimate of the
number of false detections, could result in our failure to identify
interesting new classes of objects such as quasars lensed by ``dark''
clusters (Hattori et al.\ 1997), clusters dominated by a single galaxy
(Ponman et al.\ 1994), ``failed'' clusters (Tucker, Tananbaum, \& Remillard
1995).  Therefore, it is desirable to have an independent estimate of the
number of false detections and their distribution as a function of flux and
radius. For this, we simulate \ROSAT\/ images without clusters and reduce
them identically to the real data.  All the extended sources detected in
these simulations are false. Since the simulations correctly reproduce
fluxes and spatial distribution of point sources and all the instrumental
artifacts of the \ROSAT\/ PSPC, the expected number of false detections can
be accurately measured.

Confusion of point sources is the main effect leading to false cluster
detections. The degree of confusion depends strongly on whether point
sources are distributed randomly or have angular correlation, and the number
of false detections changes correspondingly. From simulations, we derive
that our source catalog should on average contain 17.2 false detections if
point sources are randomly located. If point sources have correlation with
the observed amplitude (Vikhlinin \& Forman 1995), the number of false
detections increases to 25.9.  Fig~\ref{fig:falsedistr} shows the
distribution of false detections in radius vs.\ flux coordinates and their
cumulative distribution as a function of flux, obtained for correlated point
sources. For randomly located sources, the distributions in
Fig~\ref{fig:falsedistr} should simply be scaled. The contamination of our
extended source catalog by confused point sources is between 8\% and 11\%.

The predicted number of false detections agrees well with results of optical
identifications. From simulations, we expect on average $\approx1.5$ false
detections with fluxes $>2\times10^{-13}\,$\ergs. Of 82 X-ray sources above
this flux, 80 are optically confirmed clusters, and one is a likely false
detection. In the total sample, we expect $\approx 17-26$ false sources,
while the presently available optical identifications set an upper limit of
23 and lower limit of 18 false detections in the data
(Table~\ref{tab:optidsum}).  Finally, the distribution of flux and
core-radius of X-ray sources without optical cluster counterparts matches
well the distribution for false detections found in simulations
(Fig~\ref{fig:falsedistr}). Thus, our sample provides no support for the
existence of ``dark'' clusters.

\subsection{Sky Coverage as a Function of Flux}

To compute the $\log N - \log S$ function, the survey solid angle as a
function of flux is required. Traditionally, the sky coverage is thought of
as the area, in which a survey is ``complete'', i.e.\ all sources above the
given flux are detected. The differential $\log N - \log S$ is computed as
the ratio of the number of detected sources in a given flux bin and the sky
coverage in this flux bin. However, this view of the sky coverage is not
correct in the presence of significant flux measurement errors, which is the
case in all \ROSAT\/ surveys. First, the source detection probability
changes gradually from 0 to 1 in a flux range of finite width, and cannot be
adequately approximated by a step-like function of flux. Second, the
measurement scatter leads to significant biases in the derived $\log N -
\log S$ relation, as we describe below. Some intrinsically bright sources
have low measured fluxes, while some intrinsically faint sources have high
measured fluxes.  For surveys with uniform sensitivity, the number of
sources usually increases at faint fluxes and the described effect leads to
overestimation of $\log N - \log S$ (Eddington 1940). In X-ray surveys, the
sky coverage usually drops rapidly at faint fluxes and therefore the number
of detected sources decreases at faint fluxes. In this case, the sign of the
Eddington bias is opposite and the $\log N - \log S$ function is
underestimated (see Fig.~6 in Hasinger et al.\ 1993a).

\vskip -1ex
\tabcaption{\centerline{Sky coverage of the survey}\label{tab:area}}
\begin{center}
\vskip -0.5ex
\renewcommand{\arraystretch}{1.2}
\footnotesize
\begin{tabular}{ccc}
\hline
\hline
Limiting Flux & \multicolumn{2}{c}{Solid Angle (deg$^2$) } \\
\cline{2-3} 
\ergs  & entire sample & $z>0.5$ clusters \\
\hline
$1.3\times10^{-14}$ & 0.074 & 0.070 \\
$1.5\times10^{-14}$ & 0.094 & 0.089 \\
$2.0\times10^{-14}$ & 0.185 & 0.190 \\
$3.0\times10^{-14}$ & 1.354 & 1.364 \\
$4.5\times10^{-14}$ & 9.026 & 9.100 \\
$7.0\times10^{-14}$ & 34.74 & 34.03 \\
$1.0\times10^{-13}$ & 66.55 & 66.20 \\
$1.5\times10^{-13}$ & 102.6 & 104.3 \\
$2.0\times10^{-13}$ & 122.8 & 127.4 \\
$3.0\times10^{-13}$ & 140.9 & 147.0 \\
$4.5\times10^{-13}$ & 148.1 & 154.0 \\
$7.0\times10^{-13}$ & 149.3 & 159.6 \\
$1.0\times10^{-12}$ & 151.1 & 161.3 \\
$1.5\times10^{-12}$ & 157.1 & 164.7 \\
$2.0\times10^{-12}$ & 158.5 & 165.1 \\
\hline		  
\end{tabular}
\end{center}
\smallskip

For a plausible model of the source population, one can calculate the ratio
of the differential $\log N - \log S$ for detected and real sources, if the
detection probability and measurement scatter is known. The ratio of these
$\log N - \log S$ functions has the usual meaning of the sky coverage. This
approach to the survey area calculation was used by Vikhlinin et al.\ (1995
and 1995b) to obtain an unbiased measurement of the $\log N - \log S$
relation for point sources. We use the same approach here to define the
survey area for the present cluster survey. We assume non-evolving clusters
with the luminosity function of Ebeling et al.\ (1997) in a $q_0=0.5$
cosmology.  The distribution of cluster radii and their correlation with
luminosities is adopted from Jones \& Forman (1998). We simulate cluster
redshifts between $z=0$ and $z=2$ using the cosmological volume-per-redshift
law (e.g.\ Peebles 1993). We then simulate the rest-frame luminosity between
$L_x=10^{42}$ and $10^{46}\,$ergs~s$^{-1}$.  Cluster radius is simulated
from the distribution corresponding to the simulated luminosity.  We then
calculate the observed angular radius and flux accounting for the
correlation between X-ray luminosity and temperature (e.g., David et al.\
1993), the probability to detect this cluster (\S\ref{sec:detprob}), and
finally we simulate the measured flux (\S\ref{sec:measscat}). The detection
probability is added to the distribution of detected clusters as a function
of measured flux, and 1 is added to the the number of input clusters in the
corresponding bin of real flux. Simulating $10^6$ clusters according to this
procedure, we determine the sky coverage as the ratio of detected and input
sources in the corresponding flux bins.  The calculated survey area is shown
in Table~\ref{tab:area}. In this table we also show the sky coverage for the
distant, $z>0.5$, subsample. The sky coverage for distant subsamples differs
from that for the entire sample because of the different distribution of
angular sizes.

Using different cluster evolution models (including evolution of
luminosities, number density, and radii), we verified that the derived sky
coverage varies by no more than $10\%$ compared to the no-evolution
assumption, if cluster radii do not evolve.  Using the present cluster
sample, Vikhlinin et al.\ (1998) show that the distribution of sizes of
distant and nearby clusters is indeed very similar.

\vspace*{-1.5ex}
\centerline{\includegraphics[width=3.25in]{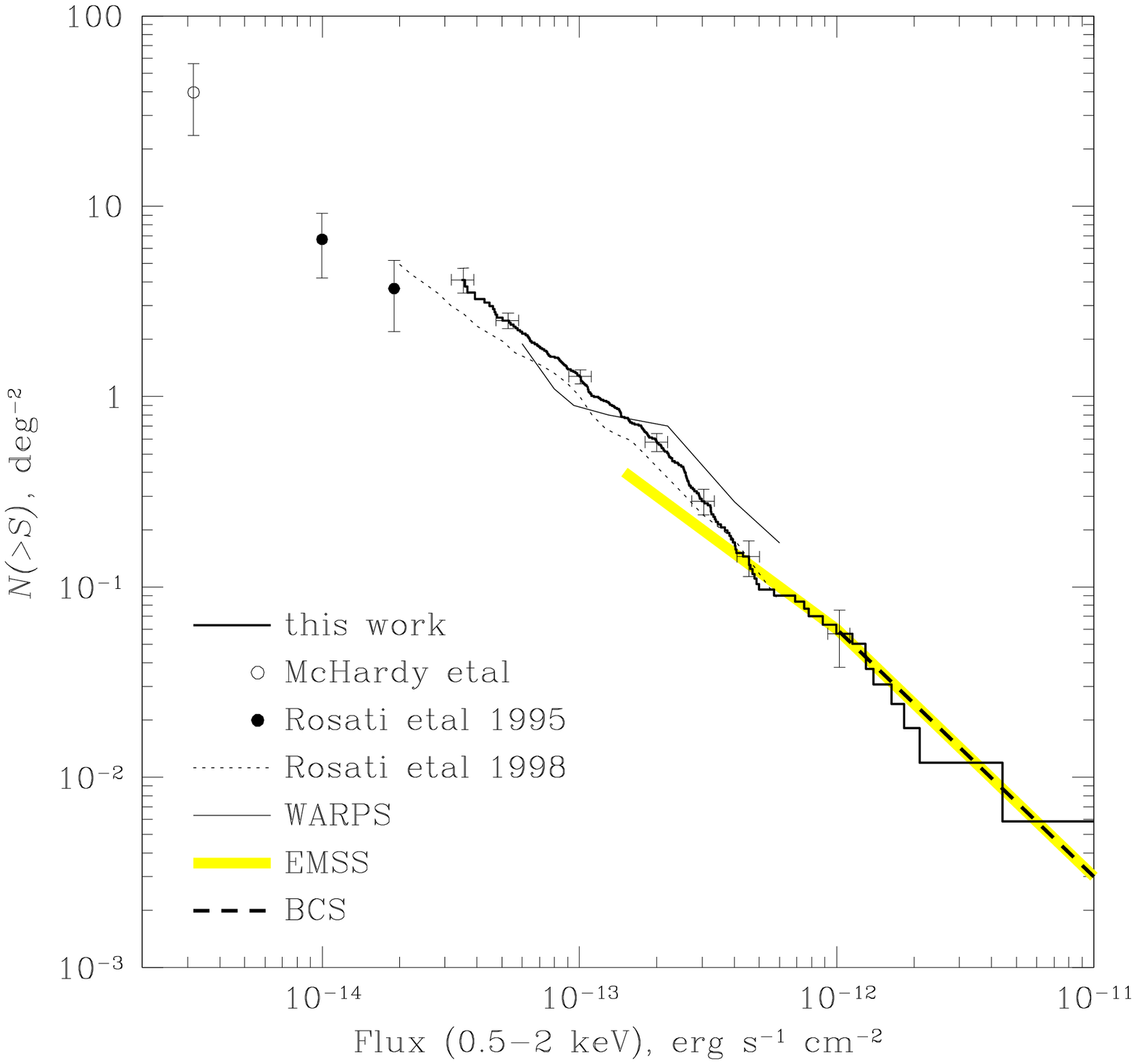}}
\vskip -2.5ex
\figcaption{Cluster $\log N - \log S$ relation. The results from
our survey are shown as the heavy solid histogram with several individual
points including error bars. Vertical error bars represent the uncertainty
in the number of clusters, while horizontal error bars correspond to a
possible systematic uncertainty in flux (\S\ref{sec:fluxcalc}).  Other
surveys are shown for comparison.\label{fig:lnls}}
\medskip

\section{${\rm LOG}\; N - {\rm LOG}\; S$~ relation for clusters}

Using the survey solid angle, we calculate the $\log N - \log S$ relation
for clusters. Each optically confirmed cluster is added to the cumulative
distribution with the weight equal to the inverse solid angle corresponding
to its measured flux. The derived cumulative $\log N - \log S$ function is
shown in Fig~\ref{fig:lnls}. We also show the cluster counts derived in
other surveys: EMSS (adopted from Jones et al.\ 1998), \ROSAT\/ All-Sky survey
sample of X-ray brightest clusters (BCS; Ebeling et al.\ 1997), WARPS survey
(Jones et al.\ 1998), Rosati et al.\ (1998 and 1995), and an ultra-deep UK
\ROSAT\/ survey (McHardy et al.\ 1997). The $\log N - \log S$ relation
derived from our survey spans more than 2.5 orders of magnitude in flux. At
the bright end, our result shows excellent agreement with the samples of
nearby clusters from the BCS and EMSS. At intermediate fluxes, around
$2\times10^{-13}\,$\ergs, our cluster counts agree well with a small-area
WARPS survey. Finally, the extrapolation of our $\log N - \log S$ relation
down to $3\times10^{-15}\,$\ergs, agrees with results of McHardy et al.\
(1997), who identified most of the X-ray sources, regardless of extent, in
their ultra-deep survey. Our $\log N - \log S$ relation seems to be
systematically higher than the surface density of clusters identified in the
50~deg$^2$ survey of Rosati et al. For example, the difference is a factor
of 1.3 at $2\times10^{-13}\,$\ergs, where we optically confirmed 98\% of
detected sources and where the survey area corrections are relatively small.
This difference is marginally significant at the $\sim 2\sigma$ level. Since
Rosati et al.\ have not published their cluster sample nor the details of
the survey area calculations, it is hard to assess the source of this
discrepancy. We only note that it can be explained, for example, if there is
a systematic difference of 15--20\% in fluxes.  A discrepancy of our $\log N
- \log S$ relation with the EMSS near their sensitivity limit is most likely
due to the difference in measured fluxes (\S\ref{sec:fluxcalc}).

\section{Summary}

We present a catalog of 200 clusters detected as extended X-ray sources in
647 \ROSAT\/ PSPC observations covering a solid angle of 158 square degrees.
To detect these sources, we used a novel detection algorithm combining a
wavelet decomposition to find candidate extended sources and Maximum
Likelihood fitting to evaluate the statistical significance of the source
extent.  Optical identifications demonstrate a high success rate of our
X-ray selection: 90\% of detected sources in the total sample, and 98\% in
the bright subsample are optically confirmed as clusters of galaxies. We
present X-ray parameters of all detected sources and spectroscopic or
photometric redshifts for optically confirmed clusters. Extensive
Monte-Carlo simulations of our source detections are used to derive the sky
coverage of the survey necessary for a statistical study of X-ray properties
of our clusters. We present the $\log N - \log S$ relation derived from our
cluster catalog.  This relation shows a general agreement with other,
smaller area surveys.

In a subsequent paper (Vikhlinin et al.\ 1998) we use this sample to
constrain the evolution of cluster luminosities and radii at high redshift.

\acknowledgements

We thank M.~Markevitch for useful discussions, D.~Fabricant and M.~Franx for
the advice regarding MMT observations, and H.~Ebeling and P.~Rosati for
useful communications regarding their surveys. We are grateful to
J.~A.~Tyson, E.~Barton, and S.~Jha who obtained some of the CCD images. We
made use of the Digitized Sky Survey produced by the Space Telescope Science
Institute from the Oschin Schmidt Telescope on Mt.\ Palomar and the UK
Schmidt Telescope photographic plates, NASA/IPAC Extragalatic Database, and
the \ROSAT\/ data archive maintained by GSFC.  Financial support for this
work was provided by the Smithsonian Institution, NAS8-39073 contract, and
the Russian Basic Research foundation grant 95--02--05933. HQ acknowledges
partial support from FONDECYT grant 8970009 and the award of a Presidential
Chair in Science.



\begin{deluxetable}{rlcrrrrclrcrc}
\tablecaption{Cluster catalog}
\tablefontsize{\footnotesize}
\tableheadfrac{0.01}
\tablewidth{0pt}
\tablehead{
\colhead{No.} &
\colhead{RA} &
\colhead{\phantom{D}Dec} &
\multicolumn{1}{c}{$F_x$} &
\multicolumn{1}{c}{$\delta F_x$} &
\multicolumn{1}{c}{$r_c$} &
\multicolumn{1}{c}{$\delta r_c$} &
\colhead{} &
\colhead{$z$} &
\colhead{$z_{\rm min}-z_{\rm max}$} &
\colhead{} &
\multicolumn{1}{r}{$\delta r$} &
\colhead{Note} \nl
\colhead{} &
\colhead{(J2000)} &
\colhead{\phantom{D}(J2000)} &
\multicolumn{2}{c}{$10^{-14}$ cgs} &
\multicolumn{1}{c}{(\arcsec)} &
\multicolumn{1}{c}{(\arcsec)} &
\colhead{} &
\colhead{} &
\colhead{} &
\colhead{} &
\multicolumn{1}{r}{(\arcsec)} &
}
\startdata
  1&  00~~30~~33.2  &  $+$26~~18~~19  &   24.3 &    3.0\phantom{X}  &  31 &   3& & 0.500 & & & 13&  \nl
  2&  00~~41~~10.3  &  $-$23~~39~~33  &    9.8 &    2.4\phantom{X}  &  25 &  12& & 0.15 & 0.08--0.22\rlap{$^{\rm dss}$}& & 19&  \nl
  3&  00~~50~~59.2  &  $-$09~~29~~12  &   36.6 &    4.9\phantom{X}  &  45 &   4& & 0.21 & 0.14--0.25& & 11&  \nl
  4&  00~~54~~02.8  &  $-$28~~23~~58  &   10.8 &    1.5\phantom{X}  &  37 &   6& & 0.25 & 0.18--0.29& & 16&  \nl
  5&  00~~56~~55.8  &  $-$22~~13~~53  &   25.9 &    5.2\phantom{X}  &  61 &  12& & 0.11 & 0.04--0.18\rlap{$^{\rm dss}$}& & 17&  \nl
  6&  00~~56~~56.1  &  $-$27~~40~~12  &    6.9 &    0.8\phantom{X}  &  14 &   2& & 0.563 & & & 13&   1\nl
  7&  00~~57~~24.2  &  $-$26~~16~~45  &  186.1 &   21.3\phantom{X}  &  82 &   6& & 0.113 & & & 14&   2\nl
  8&  01~~10~~18.0  &  $+$19~~38~~23  &    7.8 &    1.6\phantom{X}  &  35 &   8& & 0.24 & 0.17--0.28& & 16&  \nl
  9&  01~~11~~36.6  &  $-$38~~11~~12  &    8.9 &    1.7\phantom{X}  &  18 &   3& & 0.122 & & &  9&  \nl
  10&  01~~22~~35.9  &  $-$28~~32~~03  &   26.9 &    6.3\phantom{X}  &  37 &  16& & 0.24 & 0.17--0.28& & 14&   3\nl
  11&  01~~24~~35.1  &  $+$04~~00~~49  &    7.5 &    2.2\phantom{X}  &  31 &  14& & 0.27 & 0.20--0.31& & 20&  \nl
  12&  01~~27~~27.8  &  $-$43~~26~~13  &    5.7 &    1.9\phantom{X}  &  34 &  13& & \multicolumn{1}{c}{---\phantom{3}} & & & 19& F\nl
  13&  01~~28~~36.9  &  $-$43~~24~~57  &    7.5 &    1.3\phantom{X}  &  10 &   3& & 0.26 & 0.19--0.30& &  9&  \nl
  14&  01~~32~~54.7  &  $-$42~~59~~52  &   32.3 &    8.1\phantom{X}  &  75 &  25& & 0.088 & & & 23&   4\nl
  15&  01~~36~~24.2  &  $-$18~~11~~59  &    4.8 &    1.0\phantom{X}  &  21 &   8& & 0.25 & 0.18--0.29& & 15&  \nl
  16&  01~~39~~39.5  &  $+$01~~19~~27  &   10.9 &    2.0\phantom{X}  &  37 &   8& & 0.25 & 0.18--0.29& & 12&  \nl
  17&  01~~39~~54.3  &  $+$18~~10~~00  &   27.3 &    3.8\phantom{X}  &  33 &   5& & 0.176 & & &  9&   5\nl
  18&  01~~42~~50.6  &  $+$20~~25~~16  &   26.1 &    4.5\phantom{X}  &  29 &   6& & 0.43 & 0.36--0.47& & 22&  \nl
  19&  01~~44~~29.1  &  $+$02~~12~~37  &   10.1 &    2.3\phantom{X}  &  32 &  11& & 0.15 & 0.08--0.19& & 13&  \nl
  20&  01~~54~~14.8  &  $-$59~~37~~48  &   14.5 &    3.2\phantom{X}  &  22 &   7& & 0.360 & & & 12&  \nl
  21&  01~~59~~18.2  &  $+$00~~30~~12  &   32.7 &    4.1\phantom{X}  &  13 &   2& & 0.26 & 0.19--0.30& &  9&  \nl
  22&  02~~06~~23.4  &  $+$15~~11~~16  &   13.0 &    2.5\phantom{X}  &  53 &  10& & 0.27 & 0.20--0.31& & 14&  \nl
  23&  02~~06~~49.5  &  $-$13~~09~~04  &   26.0 &    4.4\phantom{X}  &  28 &   8& & 0.31 & 0.24--0.35& & 15&  \nl
  24&  02~~10~~13.8  &  $-$39~~32~~51  &    4.6 &    1.1\phantom{X}  &  22 &  10& & 0.19 & 0.12--0.23& & 11&  \nl
  25&  02~~10~~25.6  &  $-$39~~29~~47  &    6.4 &    1.3\phantom{X}  &  28 &   9& & 0.27 & 0.20--0.30& & 14&  \nl
  26&  02~~28~~13.2  &  $-$10~~05~~40  &   24.4 &    3.9\phantom{X}  &  35 &   6& & 0.149 & & & 15&  \nl
  27&  02~~36~~05.2  &  $-$52~~25~~03  &    5.8 &    1.2\phantom{X}  &  16 &   4& & 0.60 & 0.53--0.67\rlap{$^{\rm dss}$}& &  9&  \nl
  28&  02~~37~~59.2  &  $-$52~~24~~40  &   64.4 &    8.2\phantom{X}  &  49 &   8& & 0.13 & 0.06--0.20\rlap{$^{\rm dss}$}& & 14&   6\nl
  29&  02~~39~~52.6  &  $-$23~~20~~35  &    8.4 &    1.8\phantom{X}  &  51 &  14& & 0.49 & 0.42--0.53& & 23&  \nl
  30&  02~~58~~46.1  &  $+$00~~12~~44  &   10.8 &    2.9\phantom{X}  &  28 &   7& & 0.23 & 0.16--0.27& & 19&  \nl
  31&  02~~59~~33.9  &  $+$00~~13~~47  &   32.4 &    5.2\phantom{X}  &  42 &  11& & 0.17 & 0.10--0.21& & 12&  \nl
  32&  03~~22~~20.1  &  $-$49~~18~~40  &   40.3 &    7.2\phantom{X}  &  69 &  11& & 0.067 & & & 15&   7\nl
  33&  03~~37~~44.9  &  $-$25~~22~~39  &    3.7 &    0.7\phantom{X}  &   7 &   2& & 0.38 & 0.31--0.42& &  8&  \nl
  34&  03~~41~~57.1  &  $-$45~~00~~11  &    1.7 &    0.4\phantom{X}  &  27 &   9& & 0.36 & 0.29--0.43\rlap{$^{\rm dss}$}& & 12&  \nl
  35&  03~~51~~37.8  &  $-$36~~49~~50  &    8.8 &    2.2\phantom{X}  &  31 &  17& & \nodata & \nodata& & 24& f?\nl
  36&  04~~28~~43.0  &  $-$38~~05~~54  &   20.8 &    5.0\phantom{X}  &  54 &  13& & 0.154 & & & 20&   8\nl
  37&  04~~34~~15.7  &  $-$08~~31~~17  &    7.2 &    2.2\phantom{X}  &  25 &  14& & 0.24 & 0.17--0.28& & 24&  \nl
  38&  05~~05~~57.8  &  $-$28~~25~~47  &   14.2 &    1.9\phantom{X}  &  25 &   4& & 0.131 & & & 15&  \nl
  39&  05~~06~~03.7  &  $-$28~~40~~44  &   19.5 &    3.4\phantom{X}  &  84 &  20& & 0.11 & 0.04--0.18\rlap{$^{\rm dss}$}& & 21&  \nl
  40&  05~~21~~10.7  &  $-$25~~30~~44  &   17.6 &    4.0\phantom{X}  &  37 &  13& & \nodata & \nodata& & 15&  \nl
  41&  05~~22~~14.2  &  $-$36~~25~~04  &   18.4 &    3.8\phantom{X}  &  16 &   5& & 0.54 & 0.47--0.61\rlap{$^{\rm dss}$}& &  9&  \nl
  42&  05~~28~~40.3  &  $-$32~~51~~38  &   19.9 &    2.5\phantom{X}  &  26 &   3& & 0.273 & & &  8&  \nl
  43&  05~~29~~38.4  &  $-$58~~48~~10  &    5.6 &    1.0\phantom{X}  &  10 &   3& & \nodata & \nodata& &  9& f?\nl
  44&  05~~32~~43.7  &  $-$46~~14~~11  &   41.1 &    4.3\phantom{X}  &  12 &   1& & 0.10 & 0.03--0.17\rlap{$^{\rm dss}$}& &  7&  \nl
  45&  05~~33~~53.2  &  $-$57~~46~~52  &   22.2 &    6.1\phantom{X}  &  81 &  28& & 0.15 & 0.08--0.22\rlap{$^{\rm dss}$}& & 37&  \nl
  46&  05~~33~~55.9  &  $-$58~~09~~16  &    9.0 &    2.8\phantom{X}  &  53 &  20& & \nodata & \nodata& & 30& f?\nl
  47&  08~~10~~23.9  &  $+$42~~16~~24  &  238.6 &   27.2\phantom{X}  &  59 &   5& & 0.064 & & & 14&  \nl
  48&  08~~18~~57.8  &  $+$56~~54~~34  &   10.1 &    2.5\phantom{X}  &  29 &   9& & 0.260 & & & 17&  \nl
  49&  08~~19~~22.6  &  $+$70~~54~~48  &   10.1 &    1.8\phantom{X}  &  24 &   6& & 0.226 & & & 15&  \nl
  50&  08~~19~~54.4  &  $+$56~~34~~35  &   30.8 &    5.0\phantom{X}  &  16 &   5& & 0.260 & & & 14&  \nl
  51&  08~~20~~26.4  &  $+$56~~45~~22  &   22.9 &    4.2\phantom{X}  &  39 &  14& & 0.043 & & & 18&  \nl
  52&  08~~26~~06.4  &  $+$26~~25~~47  &   10.9 &    2.6\phantom{X}  &  59 &  19& & 0.351 & & & 22&  \nl
  53&  08~~26~~29.7  &  $+$31~~25~~15  &   11.1 &    4.7\phantom{X}  &  47 &  22& & 0.26 & 0.19--0.30& & 31&  \nl
  54&  08~~31~~16.0  &  $+$49~~05~~06  &   12.3 &    4.0\phantom{X}  &  30 &  15& & \multicolumn{1}{c}{---\phantom{3}} & & & 17& F\nl
  55&  08~~34~~27.4  &  $+$19~~33~~24  &    8.3 &    1.7\phantom{X}  &  31 &   7& & \multicolumn{1}{c}{---\phantom{3}} & & & 18& F\nl
  56&  08~~41~~07.4  &  $+$64~~22~~43  &   29.1 &    3.2\phantom{X}  &  35 &   3& & 0.36 & 0.29--0.40& &  8&  \nl
  57&  08~~41~~43.4  &  $+$70~~46~~53  &    8.9 &    2.1\phantom{X}  &  31 &  12& & 0.235 & & & 13&  \nl
  58&  08~~42~~52.8  &  $+$50~~23~~16  &    6.3 &    1.7\phantom{X}  &  23 &  10& & 0.48 & 0.41--0.53& & 16&  \nl
  59&  08~~47~~11.3  &  $+$34~~49~~16  &   12.2 &    3.0\phantom{X}  &  28 &   9& & 0.560 & & & 17&  \nl
  60&  08~~48~~47.6  &  $+$44~~56~~21  &    3.3 &    0.6\phantom{X}  &  14 &   4& & 0.574 & & & 13&  \nl
  61&  08~~48~~56.3  &  $+$44~~52~~16  &    2.7 &    0.6\phantom{X}  &  23 &   6& & \multicolumn{1}{c}{---\phantom{3}} & & & 14& F\nl
  62&  08~~49~~11.1  &  $+$37~~31~~25  &   14.7 &    3.0\phantom{X}  &  36 &  10& & 0.240 & & & 14&  \nl
  63&  08~~52~~33.6  &  $+$16~~18~~08  &   37.1 &    6.2\phantom{X}  &  33 &  10& & 0.098 & & & 19&  \nl
  64&  08~~53~~14.1  &  $+$57~~59~~39  &   19.8 &    5.8\phantom{X}  &  35 &  14& & 0.475 & & & 17&  \nl
  65&  08~~57~~45.7  &  $+$27~~47~~32  &    6.8 &    1.6\phantom{X}  &  42 &  11& & 0.50 & 0.43--0.54& & 27&  \nl
  66&  08~~58~~25.0  &  $+$13~~57~~16  &    6.4 &    1.0\phantom{X}  &  14 &   5& & 0.54 & 0.47--0.58& & 10&  \nl
  67&  09~~07~~17.9  &  $+$33~~30~~09  &    4.4 &    0.8\phantom{X}  &  24 &   5& & 0.46 & 0.39--0.49& & 14&  \nl
  68&  09~~07~~20.4  &  $+$16~~39~~09  &  148.5 &   17.6\phantom{X}  &  55 &   5& & 0.076 & & &  9&   9\nl
  69&  09~~10~~39.7  &  $+$42~~48~~41  &    8.3 &    2.0\phantom{X}  &  76 &  23& & \nodata & \nodata& & 24& U\nl
  70&  09~~21~~13.4  &  $+$45~~28~~50  &   23.9 &    4.7\phantom{X}  &  26 &   5& & 0.337 & & & 11&  \nl
  71&  09~~26~~36.6  &  $+$12~~42~~56  &   16.7 &    2.1\phantom{X}  &  16 &   3& & 0.50 & 0.43--0.54& &  9&  \nl
  72&  09~~26~~45.6  &  $+$12~~34~~07  &   11.7 &    3.5\phantom{X}  &  60 &  22& & \multicolumn{1}{c}{---\phantom{3}} & & & 41& F\nl
  73&  09~~43~~32.2  &  $+$16~~40~~02  &   23.1 &    3.7\phantom{X}  &  36 &   5& & 0.256 & & & 10&  \nl
  74&  09~~43~~44.7  &  $+$16~~44~~20  &   21.2 &    4.1\phantom{X}  &  69 &  13& & 0.180 & & & 17&  \nl
  75&  09~~47~~45.8  &  $+$07~~41~~18  &   13.5 &    3.7\phantom{X}  &  32 &  10& & 0.59 & 0.52--0.63& & 17&  \nl
  76&  09~~51~~47.0  &  $-$01~~28~~33  &    7.1 &    1.9\phantom{X}  &  25 &  11& & 0.53 & 0.46--0.57& & 22&  \nl
  77&  09~~52~~08.7  &  $-$01~~48~~18  &    9.3 &    2.5\phantom{X}  &  39 &  14& & \multicolumn{1}{c}{---\phantom{3}} & & & 18& F\nl
  78&  09~~53~~31.2  &  $+$47~~58~~57  &   13.0 &    5.2\phantom{X}  &  41 &  20& & \multicolumn{1}{c}{---\phantom{3}} & & & 20& F, 10\nl
  79&  09~~56~~03.4  &  $+$41~~07~~14  &   15.6 &    3.3\phantom{X}  &  13 &   6& & 0.73 & 0.66--0.77& & 13&  \nl
  80&  09~~57~~53.2  &  $+$65~~34~~30  &    9.4 &    1.7\phantom{X}  &  19 &   5& & 0.530 & & & 12&  \nl
  81&  09~~58~~13.5  &  $+$55~~16~~01  &   48.2 &    7.1\phantom{X}  &  67 &  14& & 0.20 & 0.12--0.23& & 15&   11\nl
  82&  09~~59~~27.7  &  $+$46~~33~~57  &   10.5 &    5.2\phantom{X}  &  37 &  23& & \multicolumn{1}{c}{---\phantom{3}} & & & 31& F\nl
  83&  10~~02~~40.4  &  $-$08~~08~~46  &    8.6 &    2.1\phantom{X}  &  29 &   7& & 0.62 & 0.55--0.66& & 12&  \nl
  84&  10~~10~~14.7  &  $+$54~~30~~18  &   21.0 &    2.9\phantom{X}  &  20 &   4& & 0.045 & & & 14&   12\nl
  85&  10~~11~~05.1  &  $+$53~~39~~27  &    4.7 &    1.2\phantom{X}  &  23 &   9& & 0.30 & 0.23--0.34& & 11&  \nl
  86&  10~~11~~26.0  &  $+$54~~50~~08  &   20.0 &    5.1\phantom{X}  &  94 &  22& & 0.294 & & & 24&  \nl
  87&  10~~13~~38.4  &  $+$49~~33~~07  &   45.6 &    9.8\phantom{X}  & 107 &  21& & 0.17 & 0.10--0.21& & 22&  \nl
  88&  10~~15~~08.5  &  $+$49~~31~~32  &   10.8 &    2.6\phantom{X}  &  14 &   8& & 0.45 & 0.38--0.49& & 10&  \nl
  89&  10~~33~~51.9  &  $+$57~~03~~10  &   14.5 &    4.3\phantom{X}  &  24 &   9& & 0.06 & 0.00--0.10& & 16&  \nl
  90&  10~~36~~11.3  &  $+$57~~13~~31  &   18.8 &    3.9\phantom{X}  &  15 &   6& & 0.31 & 0.24--0.35& & 13&  \nl
  91&  10~~48~~00.1  &  $-$11~~24~~07  &   18.5 &    3.6\phantom{X}  &  35 &   7& & 0.065 & & & 19&  \nl
  92&  10~~49~~02.7  &  $+$54~~24~~00  &    9.1 &    1.6\phantom{X}  &  22 &   9& & 0.20 & 0.13--0.24& & 12&  \nl
  93&  10~~53~~18.4  &  $+$57~~20~~47  &    2.5 &    0.3\phantom{X}  &  12 &   3& & 0.340 & & &  8&   13\nl
  94&  10~~56~~12.6  &  $+$49~~33~~11  &   12.9 &    1.9\phantom{X}  &  64 &  15& & 0.199 & & & 23&  \nl
  95&  10~~58~~13.0  &  $+$01~~36~~57  &  129.5 &   19.3\phantom{X}  & 113 &  13& & 0.038 & & & 15&   14\nl
  96&  11~~17~~12.0  &  $+$17~~44~~24  &   12.0 &    5.6\phantom{X}  &  65 &  33& & 0.51 & 0.44--0.55& & 26&  \nl
  97&  11~~17~~26.1  &  $+$07~~43~~35  &    6.1 &    1.6\phantom{X}  &  18 &   7& & 0.40 & 0.33--0.44& & 12&  \nl
  98&  11~~17~~30.2  &  $+$17~~44~~44  &   14.4 &    2.5\phantom{X}  &  36 &  10& & 0.63 & 0.56--0.67& & 16&  \nl
  99&  11~~19~~43.5  &  $+$21~~26~~44  &    5.5 &    0.9\phantom{X}  &  12 &   3& & 0.11 & 0.04--0.15& &  9&  \nl
  100&  11~~20~~57.9  &  $+$23~~26~~41  &   21.3 &    5.0\phantom{X}  &  29 &   8& & 0.71 & 0.64--0.75& & 16&  \nl
  101&  11~~23~~10.2  &  $+$14~~09~~44  &   18.2 &    4.9\phantom{X}  &  49 &  24& & 0.32 & 0.25--0.36& & 27&  \nl
  102&  11~~24~~03.8  &  $-$17~~00~~11  &   10.8 &    3.4\phantom{X}  &  34 &  19& & 0.41 & 0.34--0.45& & 22&  \nl
  103&  11~~24~~36.9  &  $+$41~~55~~59  &   40.1 &    9.6\phantom{X}  & 110 &  30& & 0.18 & 0.11--0.22& & 31&  \nl
  104&  11~~35~~54.5  &  $+$21~~31~~05  &   17.8 &    4.0\phantom{X}  &  72 &  20& & 0.14 & 0.07--0.18& & 17&  \nl
  105&  11~~38~~43.9  &  $+$03~~15~~38  &   15.9 &    3.7\phantom{X}  &  18 &   6& & 0.14 & 0.07--0.18& & 10&  \nl
  106&  11~~42~~04.6  &  $+$21~~44~~57  &   45.9 &   17.4\phantom{X}  &  56 &  34& & 0.18 & 0.11--0.22& & 26&  \nl
  107&  11~~46~~26.9  &  $+$28~~54~~15  &   39.2 &    5.8\phantom{X}  &  79 &  11& & 0.17 & 0.10--0.21& & 18&  \nl
  108&  11~~51~~40.3  &  $+$81~~04~~38  &    3.7 &    1.1\phantom{X}  &  27 &   7& & 0.27 & 0.20--0.31& & 14&  \nl
  109&  11~~58~~11.7  &  $+$55~~21~~45  &    4.7 &    1.0\phantom{X}  &  21 &   5& & 0.15 & 0.08--0.19& &  9&  \nl
  110&  11~~59~~51.2  &  $+$55~~31~~56  &   74.2 &    7.6\phantom{X}  &  24 &   2& & 0.081 & & &  7&   15\nl
  111&  12~~00~~49.7  &  $-$03~~27~~31  &   18.5 &    2.6\phantom{X}  &  29 &   5& & 0.39 & 0.32--0.42& & 10&  \nl
  112&  12~~04~~04.0  &  $+$28~~07~~08  &  102.6 &   11.4\phantom{X}  &  32 &   3& & 0.167 & & &  7&   16\nl
  113&  12~~04~~22.9  &  $-$03~~50~~55  &    8.7 &    1.3\phantom{X}  &  26 &   6& & 0.22 & 0.15--0.26& & 14&  \nl
  114&  12~~06~~33.5  &  $-$07~~44~~28  &  129.0 &   16.3\phantom{X}  &  64 &   7& & 0.12 & 0.05--0.16& & 15&  \nl
  115&  12~~11~~15.3  &  $+$39~~11~~38  &   26.6 &    3.8\phantom{X}  &  14 &   4& & 0.340 & & &  8&   17\nl
  116&  12~~13~~35.3  &  $+$02~~53~~26  &   14.3 &    3.0\phantom{X}  &  27 &   9& & 0.39 & 0.32--0.43& & 13&  \nl
  117&  12~~16~~19.4  &  $+$26~~33~~26  &   15.4 &    4.2\phantom{X}  &  15 &   6& & 0.428 & & & 15&  \nl
  118&  12~~18~~29.1  &  $+$30~~11~~46  &    5.3 &    1.4\phantom{X}  &  18 &   9& & 0.33 & 0.26--0.37& & 11&  \nl
  119&  12~~21~~24.5  &  $+$49~~18~~13  &   20.6 &    4.6\phantom{X}  &  34 &   8& & 0.70 & 0.64--0.74& & 18&  \nl
  120&  12~~22~~32.5  &  $+$04~~12~~02  &    6.3 &    1.6\phantom{X}  &  15 &   7& & \multicolumn{1}{c}{---\phantom{3}} & & & 12& F\nl
  121&  12~~36~~31.4  &  $+$00~~51~~43  &    4.8 &    1.2\phantom{X}  &  28 &   8& & 0.17 & 0.10--0.21& & 14&  \nl
  122&  12~~37~~25.1  &  $+$11~~41~~27  &   10.6 &    3.4\phantom{X}  &  41 &  15& & \nodata & \nodata& & 21& U\nl
  123&  12~~37~~38.6  &  $+$26~~32~~23  &    7.0 &    2.3\phantom{X}  &  31 &  12& & 0.28 & 0.21--0.33& & 14&  \nl
  124&  12~~52~~05.4  &  $-$29~~20~~46  &   21.7 &    4.2\phantom{X}  &  46 &  11& & 0.17 & 0.10--0.21& & 13&  \nl
  125&  12~~52~~11.3  &  $-$29~~14~~59  &    8.7 &    1.6\phantom{X}  &  11 &   5& & \multicolumn{1}{c}{---\phantom{3}} & & &  8& F\nl
  126&  12~~54~~38.3  &  $+$25~~45~~13  &   10.2 &    2.0\phantom{X}  &  31 &   7& & 0.193 & & & 13&  \nl
  127&  12~~54~~53.6  &  $+$25~~50~~55  &   13.2 &    2.5\phantom{X}  &  40 &   8& & 0.23 & 0.16--0.27& & 12&  \nl
  128&  12~~56~~04.9  &  $+$25~~56~~52  &    9.9 &    1.9\phantom{X}  &  30 &   7& & 0.17 & 0.10--0.21& & 11&  \nl
  129&  12~~56~~39.4  &  $+$47~~15~~19  &    5.7 &    0.8\phantom{X}  &  25 &   5& & 0.40 & 0.33--0.44& & 10&  \nl
  130&  13~~01~~43.6  &  $+$10~~59~~33  &   28.1 &    5.6\phantom{X}  &  54 &  11& & 0.30 & 0.23--0.34& & 18&  \nl
  131&  13~~09~~55.6  &  $+$32~~22~~31  &    9.0 &    2.9\phantom{X}  &  42 &  19& & 0.290 & & & 23&  \nl
  132&  13~~11~~12.8  &  $+$32~~28~~58  &   46.7 &    5.8\phantom{X}  &  22 &   3& & 0.245 & & &  8&   18\nl
  133&  13~~11~~30.2  &  $-$05~~51~~26  &   13.7 &    2.4\phantom{X}  &  36 &   6& & 0.49 & 0.42--0.53& & 20&  \nl
  134&  13~~25~~14.9  &  $+$65~~50~~29  &   10.1 &    3.1\phantom{X}  &  54 &  21& & 0.180 & & & 28&  \nl
  135&  13~~25~~43.9  &  $-$29~~43~~51  &    7.7 &    2.7\phantom{X}  &  43 &  11& & \multicolumn{1}{c}{---\phantom{3}} & & & 17& F\nl
  136&  13~~29~~27.3  &  $+$11~~43~~31  &   97.0 &   16.6\phantom{X}  & 120 &  16& & 0.023 & & & 22&   19\nl
  137&  13~~34~~31.1  &  $-$08~~22~~29  &    5.2 &    1.1\phantom{X}  &  13 &   5& & \multicolumn{1}{c}{---\phantom{3}} & & & 10& F\nl
  138&  13~~34~~34.4  &  $+$37~~56~~58  &    1.6 &    0.3\phantom{X}  &  16 &   5& & 0.308 & & & 11&   20\nl
  139&  13~~35~~03.7  &  $+$37~~50~~00  &    2.9 &    0.4\phantom{X}  &  21 &   4& & 0.382 & & &  9&   21\nl
  140&  13~~36~~42.1  &  $+$38~~37~~32  &    5.9 &    1.6\phantom{X}  &  20 &   9& & 0.180 & & & 16&  \nl
  141&  13~~37~~48.3  &  $+$48~~15~~46  &    7.1 &    1.5\phantom{X}  &  16 &   4& & \multicolumn{1}{c}{---\phantom{3}} & & & 10& F\nl
  142&  13~~37~~50.4  &  $+$26~~38~~49  &    9.6 &    2.1\phantom{X}  &  21 &   6& & 0.28 & 0.21--0.33& & 12&  \nl
  143&  13~~37~~53.3  &  $+$38~~54~~09  &   14.3 &    3.6\phantom{X}  &  32 &   9& & 0.29 & 0.22--0.33& & 17&  \nl
  144&  13~~40~~33.5  &  $+$40~~17~~47  &   16.1 &    2.5\phantom{X}  &  19 &   5& & 0.171 & & & 10&   22\nl
  145&  13~~40~~53.7  &  $+$39~~58~~11  &   34.4 &    6.9\phantom{X}  &  66 &  16& & 0.169 & & & 19&   23\nl
  146&  13~~41~~51.7  &  $+$26~~22~~54  &  814.0 &   84.6\phantom{X}  & 103 &   4& & 0.070 & & &  8&   24\nl
  147&  13~~42~~05.0  &  $+$52~~00~~37  &   12.4 &    1.9\phantom{X}  &  10 &   4& & \multicolumn{1}{c}{---\phantom{3}} & & & 13& F\nl
  148&  13~~42~~49.1  &  $+$40~~28~~11  &    7.4 &    2.0\phantom{X}  &  15 &   6& & 0.53 & 0.46--0.57& & 16&  \nl
  149&  13~~43~~25.0  &  $+$40~~53~~14  &   12.6 &    2.8\phantom{X}  &  18 &   7& & 0.140 & & & 10&  \nl
  150&  13~~43~~29.0  &  $+$55~~47~~17  &   17.5 &    2.8\phantom{X}  & 109 &  17& & 0.11 & 0.04--0.18\rlap{$^{\rm dss}$}& & 23&  \nl
  151&  13~~54~~16.9  &  $-$02~~21~~47  &   14.5 &    2.6\phantom{X}  &  27 &   4& & 0.49 & 0.42--0.53& & 11&  \nl
  152&  13~~54~~49.1  &  $+$69~~17~~20  &    6.4 &    1.9\phantom{X}  &  26 &  10& & 0.18 & 0.11--0.22& & 15&  \nl
  153&  14~~06~~16.3  &  $+$28~~30~~52  &    8.5 &    1.2\phantom{X}  &  14 &   4& & 0.546 & & &  9&  \nl
  154&  14~~06~~54.9  &  $+$28~~34~~17  &   25.7 &    3.2\phantom{X}  &  30 &   3& & 0.118 & & &  8&  \nl
  155&  14~~10~~12.4  &  $+$59~~42~~40  &   33.5 &    5.1\phantom{X}  &  38 &  12& & 0.249 & & & 18&  \nl
  156&  14~~10~~15.2  &  $+$59~~38~~31  &   20.1 &    8.7\phantom{X}  &  31 &  22& & 0.249 & & & 17&   25\nl
  157&  14~~15~~37.9  &  $+$19~~06~~33  &   25.4 &    3.4\phantom{X}  &  52 &   5& & \nodata & \nodata& & 13&   26\nl
  158&  14~~16~~28.7  &  $+$44~~46~~41  &   40.4 &    5.2\phantom{X}  &  16 &   4& & 0.400 & & &  8&  \nl
  159&  14~~18~~31.1  &  $+$25~~10~~50  &   75.6 &    7.8\phantom{X}  &  33 &   1& & 0.24 & 0.17--0.28& &  7&  \nl
  160&  14~~18~~45.2  &  $+$06~~44~~02  &   16.4 &    3.0\phantom{X}  &  18 &   5& & \multicolumn{1}{c}{---\phantom{3}} & & &  9& F\nl
  161&  14~~19~~23.5  &  $+$06~~38~~42  &   13.1 &    1.9\phantom{X}  &  17 &   4& & 0.61 & 0.54--0.65& &  9&  \nl
  162&  14~~19~~57.2  &  $+$06~~34~~26  &   10.3 &    2.1\phantom{X}  &  35 &   7& & 0.61 & 0.54--0.65& & 15&  \nl
  163&  14~~29~~38.1  &  $+$42~~34~~25  &    8.5 &    2.4\phantom{X}  &  35 &  12& & 0.30 & 0.23--0.34& & 26&  \nl
  164&  14~~38~~55.5  &  $+$64~~23~~44  &   26.2 &    3.6\phantom{X}  & 103 &  11& & \nodata & \nodata& & 19& U\nl
  165&  14~~44~~07.7  &  $+$63~~44~~58  &   17.4 &    3.2\phantom{X}  &  26 &   9& & 0.298 & & & 15&   27\nl
  166&  15~~00~~02.7  &  $+$22~~33~~51  &   14.5 &    4.5\phantom{X}  &  37 &  17& & 0.21 & 0.14--0.25& & 24&  \nl
  167&  15~~00~~51.5  &  $+$22~~44~~41  &   17.8 &    4.2\phantom{X}  &  31 &  10& & 0.450 & & & 16&  \nl
  168&  15~~15~~32.5  &  $+$43~~46~~39  &   34.6 &    9.7\phantom{X}  &  60 &  19& & 0.26 & 0.19--0.30& & 18&  \nl
  169&  15~~15~~36.8  &  $+$43~~50~~50  &   10.5 &    3.8\phantom{X}  &  34 &  18& & 0.14 & 0.07--0.18& & 22&  \nl
  170&  15~~24~~40.3  &  $+$09~~57~~39  &   30.4 &    4.1\phantom{X}  &  26 &   3& & 0.11 & 0.04--0.15& &  9&   28\nl
  171&  15~~37~~44.3  &  $+$12~~00~~26  &   26.4 &    7.4\phantom{X}  &  84 &  26& & 0.15 & 0.08--0.19& & 30&  \nl
  172&  15~~40~~53.3  &  $+$14~~45~~34  &    7.6 &    2.0\phantom{X}  &  17 &   8& & 0.45 & 0.38--0.49& & 13&  \nl
  173&  15~~44~~05.0  &  $+$53~~46~~27  &    9.7 &    2.2\phantom{X}  &  35 &  11& & 0.33 & 0.26--0.37& & 19&  \nl
  174&  15~~47~~20.7  &  $+$20~~56~~50  &   25.4 &    7.0\phantom{X}  &  51 &  20& & 0.23 & 0.17--0.28& & 24&  \nl
  175&  15~~52~~12.3  &  $+$20~~13~~45  &   49.5 &    6.0\phantom{X}  &  59 &   7& & 0.136 & & &  9&  \nl
  176&  16~~06~~42.5  &  $+$23~~29~~00  &   12.1 &    2.8\phantom{X}  &  34 &  13& & 0.310 & & & 12&  \nl
  177&  16~~20~~22.0  &  $+$17~~23~~05  &   20.8 &    3.7\phantom{X}  &  35 &   8& & 0.112 & & & 12&  \nl
  178&  16~~29~~46.1  &  $+$21~~23~~54  &   25.3 &    4.0\phantom{X}  &  46 &   8& & 0.184 & & & 19&  \nl
  179&  16~~30~~15.2  &  $+$24~~34~~59  &  179.4 &   25.9\phantom{X}  & 129 &  13& & 0.09 & 0.02--0.13& & 23&   29\nl
  180&  16~~31~~04.6  &  $+$21~~22~~02  &   29.1 &    6.4\phantom{X}  &  58 &  14& & 0.098 & & & 16&  \nl
  181&  16~~33~~40.0  &  $+$57~~14~~37  &    3.5 &    0.7\phantom{X}  &  24 &   8& & 0.239 & & & 14&  \nl
  182&  16~~39~~55.6  &  $+$53~~47~~56  &  130.5 &   14.8\phantom{X}  & 170 &   8& & 0.111 & & & 12&   30\nl
  183&  16~~41~~10.0  &  $+$82~~32~~27  &   80.5 &   10.9\phantom{X}  &  78 &  11& & 0.26 & 0.19--0.30& & 13&   31\nl
  184&  16~~41~~52.5  &  $+$40~~01~~29  &   29.4 &    7.8\phantom{X}  &  51 &  15& & 0.51 & 0.44--0.55& & 24&  \nl
  185&  16~~42~~33.5  &  $+$39~~59~~05  &    5.3 &    1.4\phantom{X}  &  19 &   9& & \multicolumn{1}{c}{---\phantom{3}} & & & 12& F, 32\nl
  186&  16~~42~~38.9  &  $+$39~~35~~53  &   10.1 &    2.3\phantom{X}  &  27 &   9& & 0.47 & 0.40--0.51& & 16&  \nl
  187&  16~~58~~34.7  &  $+$34~~30~~12  &   33.6 &    5.2\phantom{X}  &  58 &  10& & 0.330 & & & 16&  \nl
  188&  16~~59~~44.6  &  $+$34~~10~~17  &    9.8 &    3.4\phantom{X}  &  25 &  11& & 0.32 & 0.25--0.36& & 16&  \nl
  189&  17~~00~~42.3  &  $+$64~~13~~00  &   45.6 &    4.7\phantom{X}  &  18 &   1& & 0.225 & & &  7&   33\nl
  190&  17~~01~~23.0  &  $+$64~~14~~11  &   38.6 &    4.2\phantom{X}  &  25 &   2& & 0.453 & & &  7&  \nl
  191&  17~~01~~46.1  &  $+$64~~21~~15  &    3.5 &    0.7\phantom{X}  &  32 &   8& & 0.220 & & & 14&  \nl
  192&  17~~02~~13.3  &  $+$64~~20~~00  &    6.3 &    1.2\phantom{X}  &  32 &   7& & 0.224 & & & 12&  \nl
  193&  17~~22~~53.8  &  $+$41~~05~~25  &   29.4 &    6.5\phantom{X}  &  42 &  12& & 0.33 & 0.26--0.37& & 22&  \nl
  194&  17~~29~~01.9  &  $+$74~~40~~46  &   17.3 &    7.2\phantom{X}  & 100 &  31& & 0.28 & 0.21--0.35\rlap{$^{\rm dss}$}& & 40&  \nl
  195&  17~~46~~29.1  &  $+$68~~48~~54  &   22.3 &    3.2\phantom{X}  &  56 &  10& & 0.217 & & & 13&  \nl
  196&  20~~03~~28.4  &  $-$55~~56~~47  &   47.6 &    6.3\phantom{X}  &  16 &   2& & 0.015 & & &  8&   34\nl
  197&  20~~04~~49.4  &  $-$56~~03~~44  &   10.4 &    2.5\phantom{X}  &  30 &  11& & 0.71 & 0.64--0.78\rlap{$^{\rm dss}$}& & 16&  \nl
  198&  20~~05~~13.6  &  $-$56~~12~~58  &   35.0 &    4.9\phantom{X}  &   7 &   3& & \multicolumn{1}{c}{---\phantom{3}} & & &  9& F\nl
  199&  20~~59~~55.2  &  $-$42~~45~~33  &   11.2 &    1.8\phantom{X}  &   9 &   3& & 0.47 & 0.40--0.51& &  8&  \nl
  200&  21~~08~~51.2  &  $-$05~~16~~49  &   11.6 &    1.7\phantom{X}  &  34 &   7& & 0.30 & 0.23--0.34& & 12&  \nl
  201&  21~~14~~20.4  &  $-$68~~00~~56  &   25.8 &    3.3\phantom{X}  &  17 &   3& & 0.15 & 0.08--0.19& & 13&  \nl
  202&  21~~37~~06.7  &  $+$00~~26~~51  &   27.8 &    5.7\phantom{X}  &  55 &  20& & 0.05 & 0.00--0.12\rlap{$^{\rm dss}$}& & 21&   35\nl
  203&  21~~39~~58.5  &  $-$43~~05~~14  &    8.3 &    2.0\phantom{X}  &  12 &   6& & 0.30 & 0.23--0.34& & 15&  \nl
  204&  21~~46~~04.8  &  $+$04~~23~~19  &   13.8 &    2.1\phantom{X}  &  17 &   2& & 0.531 & & & 13&  \nl
  205&  22~~02~~44.9  &  $-$19~~02~~10  &    6.6 &    2.2\phantom{X}  &  36 &   9& & 0.34 & 0.27--0.38& & 22&  \nl
  206&  22~~12~~38.2  &  $-$17~~13~~55  &    5.4 &    1.4\phantom{X}  &  22 &  13& & 0.12 & 0.05--0.16& & 12&  \nl
  207&  22~~13~~31.0  &  $-$16~~56~~11  &   18.1 &    3.2\phantom{X}  &  41 &  12& & 0.32 & 0.25--0.36& & 17&  \nl
  208&  22~~39~~24.7  &  $-$05~~47~~04  &   22.2 &    3.5\phantom{X}  &  11 &   2& & 0.245 & & & 13&   36\nl
  209&  22~~39~~34.4  &  $-$06~~00~~14  &    5.9 &    2.0\phantom{X}  &  21 &  10& & 0.15 & 0.08--0.19& & 19&  \nl
  210&  22~~39~~38.9  &  $-$05~~43~~18  &   32.4 &    5.0\phantom{X}  &  34 &   5& & 0.245 & & & 15&   37\nl
  211&  22~~47~~29.1  &  $+$03~~37~~13  &   23.0 &    6.3\phantom{X}  &  46 &  17& & 0.18 & 0.11--0.22& & 20&  \nl
  212&  22~~57~~49.4  &  $+$20~~56~~25  &   11.1 &    2.1\phantom{X}  &  22 &   7& & 0.28 & 0.21--0.32& & 11&  \nl
  213&  22~~58~~07.1  &  $+$20~~55~~07  &   50.5 &    6.1\phantom{X}  &  24 &   3& & 0.288 & & &  9&   38\nl
  214&  23~~05~~26.2  &  $-$35~~46~~01  &   15.5 &    3.4\phantom{X}  &  55 &  14& & 0.21 & 0.14--0.25& & 15&  \nl
  215&  23~~05~~26.6  &  $-$51~~30~~30  &    4.2 &    1.4\phantom{X}  &  21 &  10& & 0.21 & 0.14--0.25& & 17&  \nl
  216&  23~~18~~04.8  &  $-$42~~35~~30  &   15.5 &    2.7\phantom{X}  &  28 &   8& & 0.27 & 0.20--0.31& & 17&  \nl
  217&  23~~19~~33.9  &  $+$12~~26~~17  &   38.2 &    4.7\phantom{X}  &  30 &   6& & 0.25 & 0.18--0.29& & 10&  \nl
  218&  23~~25~~39.1  &  $-$54~~43~~59  &   22.4 &    7.7\phantom{X}  &  91 &  26& & 0.10 & 0.03--0.14& & 35&  \nl
  219&  23~~28~~49.9  &  $+$14~~53~~12  &    7.6 &    1.7\phantom{X}  &  27 &  12& & 0.49 & 0.42--0.54& & 21&  \nl
  220&  23~~31~~52.1  &  $-$37~~47~~11  &   10.8 &    4.7\phantom{X}  &  46 &  25& & 0.26 & 0.18--0.29& & 28&  \nl
  221&  23~~48~~53.7  &  $-$31~~17~~20  &   32.5 &    5.1\phantom{X}  &  43 &   8& & 0.21 & 0.14--0.28\rlap{$^{\rm dss}$}& & 12&   39\nl
  222&  23~~49~~07.6  &  $-$31~~22~~26  &    6.0 &    1.4\phantom{X}  &  21 &   6& & \multicolumn{1}{c}{---\phantom{3}} & & & 11& F\nl
  223&  23~~55~~11.8  &  $-$15~~00~~26  &   26.6 &    6.7\phantom{X}  &  70 &  20& & 0.15 & 0.08--0.19& & 26&  \nl
\enddata
\tablecomments{
{\it  1}---   J1888.16 cluster  $z=0.563$                                                             
{\it  2}---   A122 $z=0.11278$                                                                        
{\it  3}---   Abell S154                                                                              
{\it  4}---   APMBGC 244-064-098 $z=0.08764$                                                          
{\it  5}---   A277 $z=0.17625$                                                                        
{\it  6}---   A3038                                                                                   
{\it  7}---   Abell S346 $z=0.067$                                                                    
{\it  8}---   A3259                                                                                   
{\it  9}---   A744 $z=0.0756$                                                                         
{\it 10}---   2\arcmin\ of PDCS 040                                                                   
{\it 11}---   A899                                                                                    
{\it 12}---   J101016.1+543006 group $z=0.045$                                                        
{\it 13}---   $z=0.340$ cluster (Schmidt et al.\ 1998)                                                
{\it 14}---   UGC 06057 group $z=0.0382$                                                              
{\it 15}---   MS 1157.3+5548 $z=0.081$                                                                
{\it 16}---   MS 1201.5+2824 $z=0.167$                                                                
{\it 17}---   MS 1208.7+3928 $z=0.340$                                                                
{\it 18}---   MS 1308.8+3244 $z=0.245$                                                                
{\it 19}---   MKW 11 $z=0.02314$                                                                      
{\it 20}---   $z=0.308$ cluster (McHardy et al.\ 1997)                                                
{\it 21}---   $z=0.382$ cluster (McHardy et al.\ 1997)                                                
{\it 22}---   RX J13406+4018 group $z=0.171$                                                          
{\it 23}---   A1774 $z=0.1691$                                                                        
{\it 24}---   A1775 $z=0.0696$                                                                        
{\it 25}---   probably part of A1877 $z=0.2493$                                                       
{\it 26}---   Image saturated by a nearby star                                                        
{\it 27}---   A1969 $z=0.29809$                                                                       
{\it 28}---   Distant cluster behind the nearby group                                                 
{\it 29}---   MCG $+04-39-010$ group                                                                  
{\it 30}---   A2220 $z=0.1106$                                                                        
{\it 31}---   TTR95 1646+82 cluster                                                                   
{\it 32}---   QSO 1640+400 $z=1.59$                                                                   
{\it 33}---   A2246 $z=0.225$                                                                         
{\it 34}---   Abell S840 $z=0.0152$                                                                   
{\it 35}---   UGC 11780 group                                                                         
{\it 36}---   part of A2465, $z$ from Jones et al.\ (1995)                                            
{\it 37}---   part of A2465, $z$ from Jones et al.\ (1995)                                            
{\it 38}---   Zw2255.5+2041 $z=0.288$                                                                 
{\it 39}---   A4043                                                            }
\label{tab:catalog}
\end{deluxetable}

\clearpage
\begin{center}
\tabcaption{\centerline{List of \ROSAT\/ pointings}\label{tab:lofields}}

\includegraphics{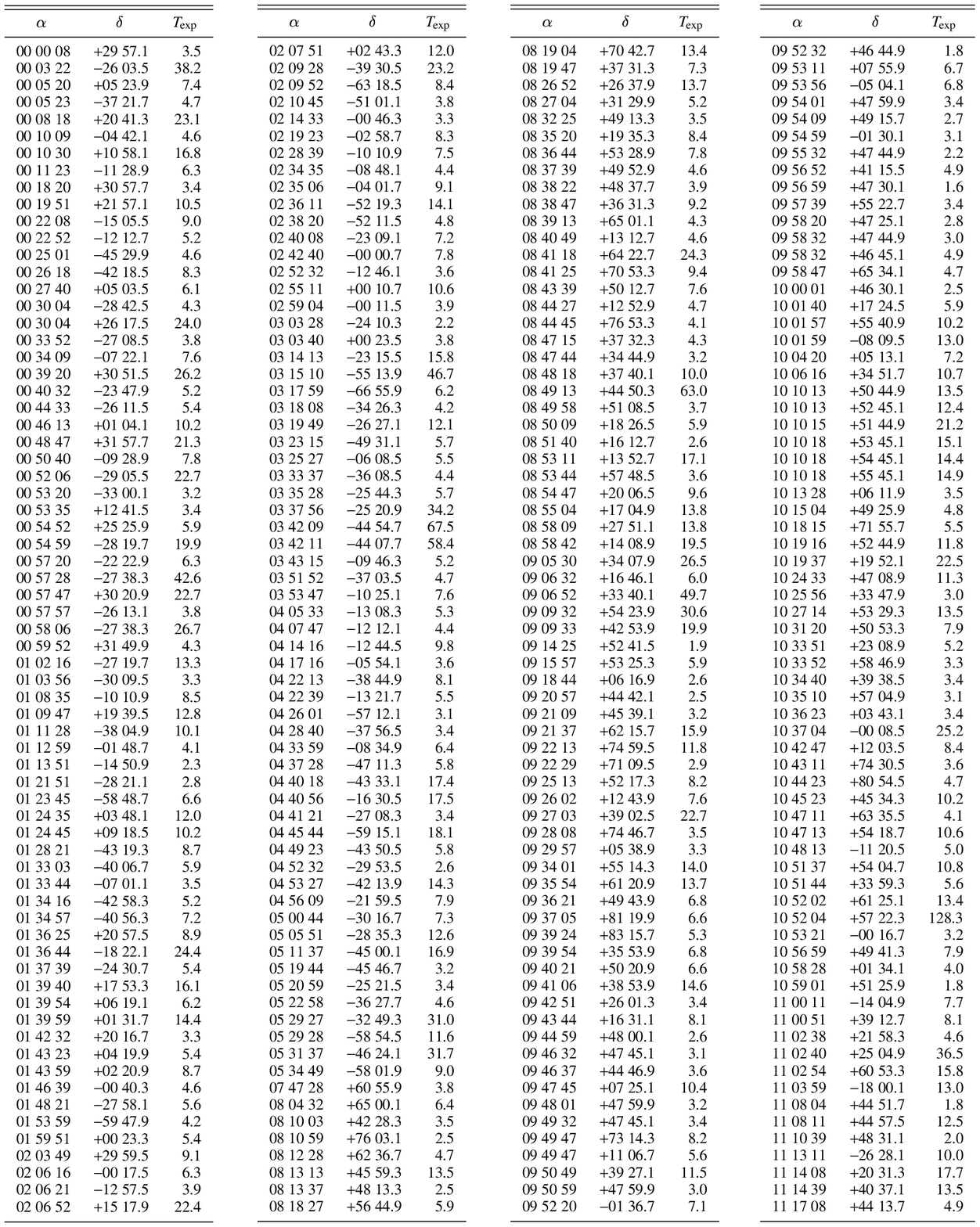}

\clearpage
\centerline{{\sc Table~\thetable}---{\it Continued}}
\vskip1.5ex
\includegraphics{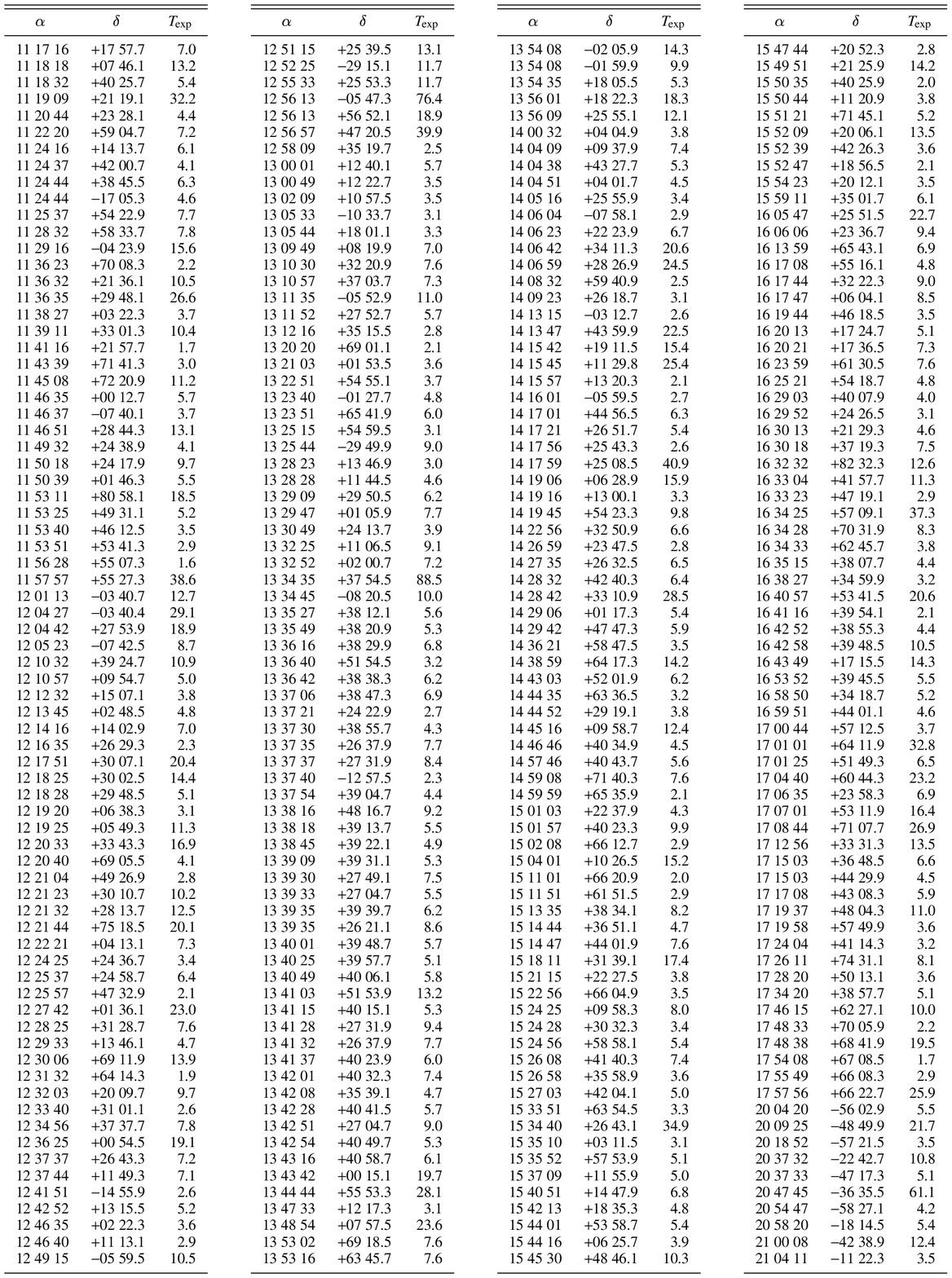}
\clearpage
\centerline{{\sc Table~\thetable}---{\it Continued}}
\vskip1.5ex
\includegraphics{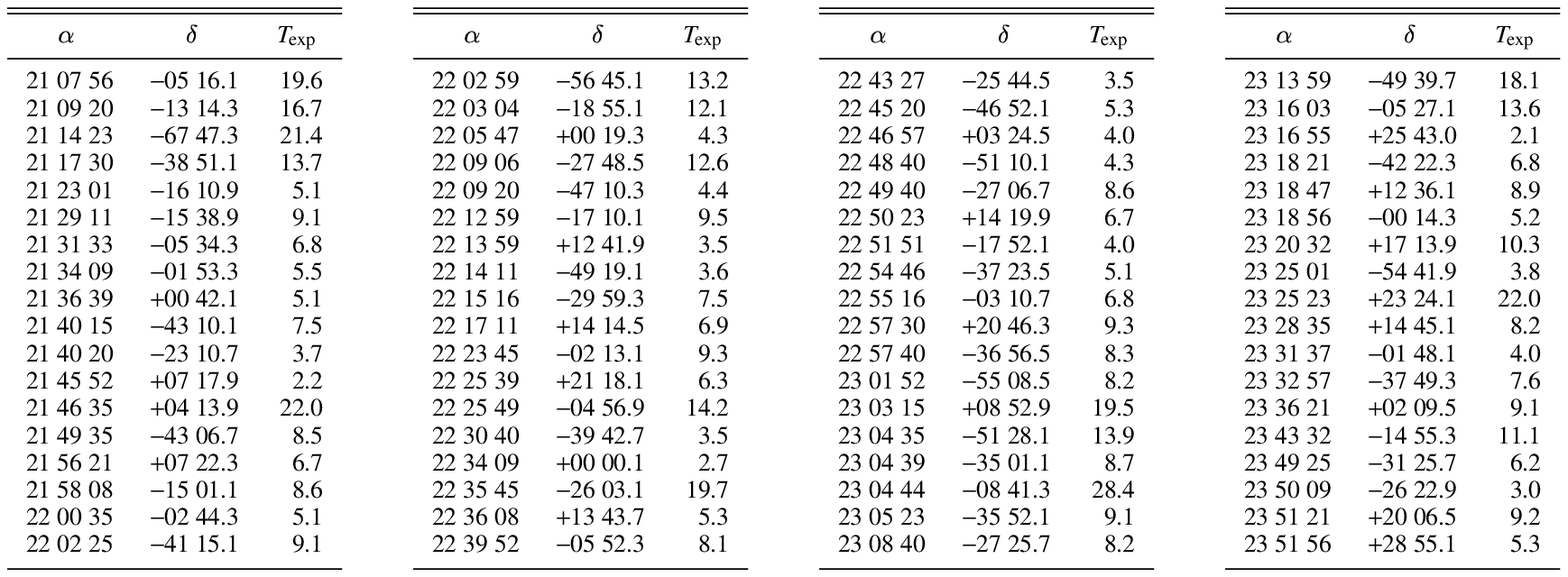}

\end{center}

\end{document}